\documentclass[capt,preprint]{aastex}
\let\new=\newcommand
\new{\be}{\begin{equation}}
\new{\ee}{\end{equation}}
\new{\disp}{\displaystyle}
\new{\bdm}{\begin{displaymath}}
\new{\edm}{\end{displaymath}}
\new{\bi}{\begin{itemize}}
\new{\ei}{\end{itemize}}
\new{\ben}{\begin{enumerate}} 
\new{\een}{\end{enumerate}}
\new{\bfig}{\begin{figure}}
\new{\efig}{\end{figure}}
\new{\bea}{\begin{eqnarray*}}
\new{\eea}{\end{eqnarray*}}

\usepackage{amsmath,amssymb,graphics,graphicx,verbatim,rotating,lscape,enumerate,enumitem}
\shorttitle{Separable solutions of force-free spheres and applications}
\shortauthors{Prasad, Mangalam \& Ravindra}
\begin{document}
\title{\uppercase{Separable solutions of force-free spheres and applications to solar active regions}}
\author{A. Prasad, A. Mangalam and B. Ravindra}
\affil{Indian Institute of Astrophysics, \\
Sarjapur Road, Koramangala 2nd Block, Bangalore 560034, India}
\email{avijeet@iiap.res.in, mangalam@iiap.res.in, ravindra@iiap.res.in}
\begin{abstract}
In this paper, we present a systematic study of the force-free field equation for simple
axisymmetric configurations in spherical geometry and apply it to the solar active regions. 
The condition of separability 
of solutions in the radial and angular variables leads to two classes of solutions:
linear and nonlinear force-free fields. 
We have studied these linear solutions 
and extended the nonlinear solutions for the radial power law index to the irreducible rational
form $n= p/q$, which is allowed for all cases of odd $p$ and cases of $q>p$ for even $p$,
where the poloidal flux $\psi\propto1/r^n$ and field $\mathbf{B}\propto 1/r^{n+2}$).
We apply these solutions to simulate photospheric vector magnetograms obtained 
using the spectropolarimeter on board \textit{Hinode}.
The effectiveness of our search strategy is first 
demonstrated on test inputs of dipolar, axisymmetric,
 and non axisymmetric linear force-free fields.
 Using the best-fit to these magnetograms, we build 
three-dimensional axisymmetric field configurations and calculate the
energy and relative helicity 
with two independent methods, which are in agreement.
We have analyzed five magnetograms for AR 10930 spanning 
a period of three days during which two X-class flares occurred which allowed us to 
find the free energy and relative helicity of the active region
before and after the flare; our analysis indicates a peak in
 these quantities  before the flare events which is consistent with 
the results mentioned in literature. We also analyzed single-polarity regions AR 10923 and 10933,
which showed very good fits with potential fields.
This method can provide useful reconstruction of the nonlinear force-free (NLFF) fields 
as well as reasonably good input fields for other numerical techniques.
\end{abstract}


\section{\uppercase{Introduction}}
The active regions in the solar photosphere are locations of high magnetic 
field where magnetic pressure starts to 
dominate over gas pressure. In such conditions the plasma is likely to follow a 
force-free equation of state, where the Lorentz force vanishes at all points.
It was shown by \citet{taylor} that in systems where 
magnetic forces are dominant in the presence of kinematic viscosity, linear force-free
fields are natural end configurations. A more general class of force-free fields is obtained when the energy of the system is
minimized with constraints of total mass, angular momentum,
cross helicity and relative helicity (e.g., \citet{finn83}; \citet{mangalam}). 
Within the context of force-free configurations, there are numerous
possibilities that can be obtained due to underlying geometry and
symmetry of the problem in addition to the invariants involved. 
There have been several attempts to construct such full three-dimensional
(3D) models from two-dimensional (2D) data obtained at one level 
vector magnetograms. A summary of the various numerical techniques are
discussed in \citet{schrijver06} and \citet{metcalf08}. They compare six algorithms for
the computation of nonlinear force-free (NLFF) magnetic
fields, which include optimization \citep{wheat00,wie04,wie06}, magnetofrictional \citep{yang86,myc94,
 roume,myc97}, Grad–Rubin based \citep{grad,amari97,amari06,wheat07,wheat09,wheat10}, and
Green’s function-based methods \citep{yan97,yan00,yan2005,yan2006} by evaluating their performance in tests on analytical
force-free field models for which boundary conditions are specified either for the entire surface
area of a cubic volume or for an extended lower boundary. Figures of merit were used to compare
the input vector field to the resulting model fields. Based on these, they argue that
all algorithms yield NLFF fields that agree best with the input field in the lower central region of the
volume, where the field and electrical currents are strongest and the effects of boundary conditions the weakest.
 The NLFF codes when applied to solar data,
do not necessarily converge to a single solution. To address
this \citet{wheat11} include uncertainties on the electric current densities
 at the boundaries iteratively until the two nonlinear solutions agree, leading
to a more reliable construction.

Because the NLFF techniques require good input fields for fast convergence and
are subject to uncertainties at the boundary conditions that propagate during 
extrapolation, we are exploring fits of the data directly to analytic solutions. 
The best-fit to a well-known (non)linear (semi)analytic solution would give us more insight into the kind of structure
that could be present in the volume given an optimal correlation with the fields observed on the magnetogram. 
The solution thus found can then be exploited to yield quantities of 
interest such as relative helicity and free energy that can be computed for the 3D configuration. Further, one can explore the stability and dynamics of these structures at a later stage.

Whereas there are several possible topologies for various geometries and boundary conditions,
e.g., \citet{marsh96}, it is our goal here to take the simplest geometric approach of a sphere. 
We show that separability condition leads to two  classes of solutions:
 linear and nonlinear force-free fields.
We call these linear fields as Chandrasekhar solution \citep{chandra56}, 
hereafter referred to as C modes and the nonlinear fields as Low-Lou  
solutions \citep{low90}, hereafter referred to as LL modes.
These computationally cheap  3D analytic models are comparable with other numerics or
with observations and this allows us to make
more precise predictions of the physically relevant configurations.
Because the validity of physical assumptions can vary from active
region to active region, we restrict ourselves to exploring the most
simplest of solutions involving the least number of parameters, namely
the choice of the modes and the two of the three Euler angles that will
represent any arbitrary rotation of the configuration space into
the coordinates of the observed magnetogram. An outline of this approach was
previously presented in \citet{prasad13}.

The paper is presented as follows:
In Section \ref{shell} we describe the formulation of the free energy and 
relative helicity in a shell geometry.
In Section \ref{modes}, we show that the force-free field equation under assumption
of axisymmetry leads to linear (C modes) and nonlinear (LL modes) force-free fields which are
discussed in Section \ref {csol} and Section \ref{llsol}, respectively. In Section \ref{s:simulation},
 we present the construction of magnetogram templates and the search strategy for obtaining the best-fit using suitable 
fitting parameters. In Section \ref{s:prepare} and Section \ref{compres} we present the data
 used for this study and compare them with the simulated models. The summary and conclusions are presented in Section \ref{s:summary}.
Details of mathematical derivations for some of the relations are are in the Appendices A-H.
Table \ref{t:formulae} provides a formulary for the C and LL modes.

\section{\uppercase{Formulation of the free energy and relative helicity in shell geometry}}
\label{shell}

In this paper, we study the solutions of axisymmetric linear and NLFFs
 in a spherical shell geometry and calculate the relevant quantities like
 free energy and relative helicity for these configurations.
The free energy of the system is the difference between the energies of a force-free 
field  and a potential field  in the entire volume.
The expression for free energy $E_{\mathrm{free}}$ is given by
\begin{equation}
 E_{\mathrm{free}}=E_{\mathrm{ff}}-E_P,
\label{efree}
\end{equation}
where $E_{\mathrm{ff}}$ and $E_P$ are the energies of the force-free field and the potential field,
respectively. Because the potential field is the minimum energy configuration for a given boundary 
condition, $E_{\textrm{free}}$ is always positive. 
Relative helicity is a gauge-invariant measure of linkages between the field lines
 with respect to a potential field matching to the perpendicular field at the surface \citep{berger}.
 Relative helicity can be computed using the Finn\textendash Antonsen formula \citep{finn85}
\begin{equation}
 H_{\mathrm{rel}}=\int_V (\mathbf{A}+\mathbf{A}_P)\cdot(\mathbf{B}-\mathbf{B}_P)dV,
\label{hrel}
\end{equation}
where $\mathbf{A}_P$ and $\mathbf{B}_P$ are the vector potential and magnetic field for the 
potential field with the constraint that $(\mathbf{B}_P)_r$ = $\mathbf{B}_r$, where $r$ represents
the radius at the boundary.
Another expression that can be used for calculating relative helicity in
spherical geometry that is independent of the potential field follows the treatment given 
in \citet{berger85}, where 
\begin{equation}
 H_{\mathrm{rel}}=2\int_V \mathbf{L}P\cdot\mathbf{L}T dV,
\label{hrelbpt}
\end{equation}
 $\mathbf{L}=-\mathbf{r}\times\nabla$ is the angular momentum operator, and $P$ and $T$ are the 
poloidal and toroidal components of the magnetic field respectively.
 The expression in Equation (\ref{hrelbpt}) can be further simplified  
 for axisymmetric magnetic fields in spherical geometry. The toroidal component
$\mathbf{L}T=B_\phi \hat{\phi}$, whereas $\mathbf{L}P=A_\phi \hat{\phi}+\nabla \psi$, which includes 
the gauge term $\nabla \psi$; $A_\phi$ and $B_\phi$ are the $\phi$ components of the vector potential 
and the magnetic field. We now use the gauge invariance of Equation (\ref{hrelbpt}) to get the final 
expression for relative helicity to be
\begin{equation}
 \int_V\mathbf{L}P\cdot\mathbf{L}T dV=\int_V A_\phi B_\phi dV+
\int_V\nabla \psi\cdot (B_\phi \hat{\phi}) dV.
\label{bergsimp}
\end{equation}
The last term in the right-hand side of Equation (\ref{bergsimp}) vanishes as
\begin{equation}
 \int_V\nabla \psi\cdot (B_\phi \hat{\phi}) dV=
\int_V\nabla\cdot(\psi B_\phi\hat{\phi})dV
=\int_S(\psi B_\phi)\hat{\phi}.\hat{r} dS =0.
\end{equation}
Thus Equation (\ref{bergsimp}) simplifies to
\begin{equation}
 H_{\mathrm{rel}}=2\int_V A_\phi B_\phi dV.
\label{hrelb}
\end{equation}
In the above derivation, it is seen that $\mathbf{L}P$ and $\mathbf{L}T$ are parallel to each other and
perpendicular to the surface normal, which leads to $H_{rel}$ being independent of the choice of $\psi$.
 This is peculiar only to certain geometries like spherical and planar.
Also see \citet{low2006}, where an absolute helicity is derived independent of the 
potential field in the special geometries that employ Euler potentials.
In the case of the linear models used here (C modes; Section \ref{csol}) the energy and helicity are finite and
 in the case of the nonlinear model used here (LL modes; Section \ref{llsol}),
$\disp \mathbf{B} \propto r^{-n-2}$ ($n>1$) and energy and helicity show singular behavior 
at the origin.

\section{\uppercase{Axisymmetric separable linear and nonlinear force-free fields}}
\label{modes}

The force-free magnetic field $\mathbf{B}$ is described by the equation
\begin{equation}
\mathbf{\mathbf{\nabla}}\times\mathbf{B}=\alpha\mathbf{B}
\label{curlb}
\end{equation}
from which it follows that $\displaystyle \mathbf{B}\cdot\mathbf{\mathbf{\nabla}}\alpha=0\label{gradal}$.
This requires $\alpha$ to be a constant along the 
magnetic field lines. Following the treatment in \citet{low90}, we assume an axisymmetric magnetic field configuration
and express it in terms of two scalar functions $\psi$ and $Q(\psi)$ in spherical polar coordinates:
\begin{equation}
 \mathbf{B}= \frac{1}{r \sin\theta}\left(\frac{1}{r}\frac{\partial \psi}{\partial\theta}\hat{\mathbf{r}}-
\frac{\partial \psi}{\partial r}\hat{\boldsymbol{\theta}}+Q\hat{\boldsymbol{\phi}}\right),
\label{spb}
\end{equation}
which is divergence-free by construction.
For an orthonormal coordinate system with a metric defined as $ds^2=g_{ii}dx^idx^i$,
the line element along the magnetic field line $dl$ is given by
$\disp {\mathbf{\hat{l}}=\sqrt{g_{ii}}\frac{dx^i}{ds}\hat{i}=\frac{B_i}{|\mathbf{B}|}\hat{i}}$;
hence $\disp \frac{\sqrt{g_{ii}}dx^i}{B_i}$ represents the equation for lines of force, and applying this 
in axisymmetry gives $\disp \psi=$ const., whose contours  represent the poloidal field lines.
Combining the Equations \ref{curlb} and \ref{spb}, we obtain
\begin{equation}
\alpha=\frac{dQ}{d\psi}\label{alp}
\end{equation}
and
\begin{equation}
 \frac{\partial^2 \psi}{\partial r^2}+\frac{(1-\mu^2)}{r^2}\frac{\partial^2 \psi}{\partial \mu^2}+Q \frac{dQ}{d\psi}=0,
\label{diff}
\end{equation}
where $\mu=\cos\theta$. To solve the above equation we choose a separable form of the type
\begin{equation}
 \psi=f(r)P(\mu),\quad Q=a \psi^\beta,
\label{sep}
\end{equation}
where $f$ and $P$ are scalar functions of $r$ and $\mu$, respectively; 
$a$ and $\beta$ are constants. Combining Equations (\ref{diff}) and (\ref{sep}), it follows that
\begin{equation}
 r^2\frac{f^{\prime\prime}}{f}+(1-\mu^2)\frac{P^{\prime\prime}}{P}+a^2\beta r^2 f^{2\beta-2}P^{2\beta-2}=0.\label{fpsep}
\end{equation}
 The first term in the left-hand side of the above equation is a function of $r$ alone and the second term is that of $\mu$
 alone. The resulting two possibilities for obtaining separable solutions are that the third term be a
 function of either
\begin{enumerate}[label=(\alph*)]
 \item $r$ alone, which is satisfied if $\beta=1$; these solutions were presented in Chandrasekhar
(1956), and we refer to them as C modes, or
\item $\mu$ alone, which is satisfied if $r^2f^{2\beta-2}=1$; these solutions were partially
explored by Low \& Lou (1990) and are termed here as LL modes.
\end{enumerate}

\section{The $\beta=1$ case: C modes }
\label{csol}
The C modes pertain to the linear force-free fields because the condition $\beta=1$ 
along with Equation (\ref{alp}) imply $\alpha=a$, and it follows from Equation (\ref{fpsep}) that
\begin{equation}
 r^2\frac{f^{\prime\prime}}{f}+a^2 r^2+(1-\mu^2)\frac{P^{\prime\prime}}{P} =0.\label{linfp}
\end{equation}
The radial part of the above equation is given by
\begin{equation}
 r^2\frac{f^{\prime\prime}}{f}+a^2 r^2=n'(n'+1)
\label{besseq}
\end{equation}
where $n'$ is a constant whose solutions are
\begin{equation}
 f_{n'}(r)=c_1\sqrt{r}J_{(1+2n')/2,}(ar) + c_2\sqrt{r}Y_{(1+2n')/2}(ar)
\end{equation}
where $J$ and $Y$ are cylindrical Bessel functions; $c_1$ and $c_2$ are
constants to be determined from the boundary conditions. The angular part of Equation (\ref{linfp}) is given by
\begin{equation}
 (1-\mu^2)\frac{P^{\prime\prime}}{P} =-n'(n'+1),
\end{equation}
whose solution is given by
\begin{equation}
 P(\mu)= (1-\mu^2)^{1/2}P_{n'}^1(\mu),
\end{equation}
where $P_{n'}^1$ is the associated Legendre function of the first kind for integer $n'$.
This solution is equivalent to that obtained in \citet{chandra56}, and the following equations give 
the correspondence between the solutions
\begin{eqnarray}
a&\leftrightarrow&\alpha\nonumber\\
f_{n'}(r)&\leftrightarrow&\sqrt{r}g_{m+3/2}(\alpha r)\nonumber\\
P_{n'}^1(\mu)&\leftrightarrow&-(1-\mu^2)^{1/2} C_m^{3/2}(\mu)\nonumber\\
n'&\leftrightarrow&m+1
\end{eqnarray}
where $C_m^{3/2}(\mu)$ denotes the Gegenbauer polynomial and $g_{m+3/2}(\alpha r)$ represents any arbitrary
linear combination of the cylindrical Bessel functions $J_{m+3/2}(\alpha r)$ and $Y_{m+3/2}(\alpha r)$.
Henceforth for the calculations of C modes we will be using the expressions from \citet{chandra56}.
Now $\psi$ can be rewritten as
\be
\psi= f(r) P(\mu)=f_{n'}(r) (1-\mu^2)^{1/2} P_{n'}^1(\mu)= r^2 S_m(r) (1-\mu^2),
\ee
where 
\begin{equation}
 S_m=\frac{g_{m+3/2}(\alpha r)}{r^{3/2}}C_m^{3/2}(\mu).
\label{sn}
\end{equation}
The application of these solutions to the case of finite spheres
 under suitable boundary conditions is discussed in Section\ref{cmode}.
The above expression can be further simplified by substituting for $S_m$ using 
Equations (\ref{sn}) and (\ref{far}) to
\begin{eqnarray}
\mathbf{B}&=&\Bigl(\frac{-J_{m+3/2}(\alpha r)}{r^{3/2}}\frac{d }
{d \mu}[(1-\mu^2)C_m^{3/2}(\mu)],\nonumber\\
&&\frac{-1}{r}\frac{d }
{d r}[r^{1/2}J_{m+3/2}(\alpha r)](1-\mu^2)^{1/2}C_m^{3/2}(\mu),
\frac{\alpha J_{m+3/2}(\alpha r)}{r^{1/2}}(1-\mu^2)^{1/2}C_m^{3/2}(\mu)\Bigr).
\label{bchand}
\end{eqnarray}
The derivation for the potential field corresponding to Equation (\ref{bchand}) is given in Appendix \ref{pot}. The final expressions
for the potential field is found to be 
\begin{equation}
\mathbf{B}_P=\left(\left[(m+1) a_{m+1} r^{m}-\frac{(m+2)b_{m+1}}{r^{m+3}}\right]P_{m+1}(\mu),-(1-\mu^2)^{1/2}
\left[ a_{m+1} r^{m}+\frac{b_{m+1}}{r^{m+3}}\right]\frac{dP_{m+1}}{d\mu},0\right).
\label{bpchand1}
\end{equation}
 $P_{(m+1)}(\mu)$ are the Legendre polynomials, where the coefficients are calculated to be
\begin{eqnarray}
\chi_l=\chi_{m+1}(r_1)&=&\frac{(m+1)(m+2)}{r_1^{3/2}}J(m+3/2,\alpha r_1)\\
a_l=a_{m+1}&=&\frac{\chi_{m+1}(r_1)}{(m+1)}\frac{r_1^{m+3}}{ r_1^{2m+3}-r_2^{2m+3}}\nonumber\\
b_l=b_{m+1}&=&\frac{(m+1)}{(m+2)}a_{m+1} r_2^{(2m+3)}
\end{eqnarray}

For the general case of open field lines, where the field has a nonzero normal component at the 
boundaries, the energy of the force-free field is given by 
\begin{equation}
 E_{\mathrm{ff}}(\alpha,n,m, r_1, r_2)=\frac{1}{4}\int_{r_1}^{r_2}\int_{-1}^{1}(B_r^2+B_\theta^2+B_\phi^2)r^2drd\mu.
\label{eint1}
\end{equation}
Upon evaluation, the above equation takes the following form

\begin{eqnarray}
E_{\mathrm{ff}}(\alpha,n,m, r_1, r_2)&=&\frac{(m+1)(m+2)}{2(2m+3)}\Bigl 
 [2\int_{r_1}^{r_2}\alpha^2r J^2_{m+3/2}(\alpha r)dr\nonumber\\
&-&r_1^{1/2} J_{m+3/2}(\alpha r_1)\frac{d}
{d r}\{r^{1/2}J_{m+3/2}(\alpha r)\}|_{r=r_1}\Bigr]
  \label{eint}
  \end{eqnarray}
An alternative and equivalent expression for the energy can also be obtained from Equation 
(\ref{eviri}), from 
which  $E_{\mathrm{ff}}= E_v(r_2)-E_v(r_1)$, where
\begin{equation}
E_v(r)=\frac{(m+1)(m+2)}{2(2m+3)}\left [r\left [\frac{d}{d r}
\left\{r^{1/2} J_{m+3/2}(\alpha r) \right\}\right]^2+\left\{\alpha^2r^2-(m+1)(m+2)
\right\}J^2_{m+3/2}(\alpha r)\right].
\label{evcmode}
 \end{equation}
We have verified that Equations (\ref{evcmode}) and (\ref{eint}) are analytically equivalent through the use of
Equation (\ref{besseq}) and algebraic manipulation of Bessel identities.
In order to calculate the free energy of the configuration using Equation (\ref{efree}), we compute 
the energy of the potential field constructed from this force-free field (see Appendix \ref{cpot})
 which is given by
\begin{eqnarray}
E_{pot}(m,r_1,r_2)&=&\frac{1}{2(2m+3)}\int_{r_1}^{r_2}\Bigl[\left((m+1)a_{m+1}r^{m+1}-\frac{(m+2)b_{m+1}}{r^{m+2}}\right)^2\nonumber\\
&+&(m+1)(m+2)\left(a_{m+1}r^{m+1}+\frac{b_{m+1}}{r^{m+2}}\right)^2\Bigr]dr.
\label{epotchand}
\end{eqnarray}
We now calculate the relative helicity of the configuration using Equation (\ref{hrel}). The relevant quantities
to be calculated for this purpose are the vector potentials for the force-free field $\mathbf{A}$ and that of the potential
field $\mathbf{A_P}$. We use gauge freedom for the vector potential to write
$\disp \mathbf{A}=\mathbf{B}/\alpha$. The vector potential for the potential field is calculated 
in Appendix \ref{vpot}, and is given by
\begin{equation}
\mathbf{A}_P=\left(0,0,(1-\mu^2)^{1/2} P^\prime_{m+1}(\mu)\left[\frac{ a_{m+1} r^{m+1}}{m+2}
-\frac{b_{m+1}}{(m+1)r^{m+2}}\right]\right).
\label{Apchand1}
\end{equation}
The relative helicity for region $II$ can be written as
\begin{eqnarray}
 &&H_{rel}(\alpha,n,m,r_1,r_2)=\int\left(\frac{B^2}{\alpha}+\mathbf{A}_P\cdot\mathbf{B}-\frac{\mathbf{B}}{\alpha}\cdot\mathbf{B}_P\right)dV\nonumber\\
 &&=\frac{8\pi E_{\mathrm{ff}}}{\alpha}+\frac{4\pi(m+1)(m+2)}{\alpha (2m+3)}\Bigl[\alpha^2 
\int_{r_1}^{r_2}\left(\frac{a_{m+1}r^{m+1}}{m+2} -\frac{b_{m+1}}{(m+1) r^{m+2}}\right )r^{3/2}J_{m+3/2}
(\alpha r) d r \nonumber\\
&&+r_1^{1/2}\left(a_{m+1}r_1^{m+1}+\frac{b_{m+1}}{r_1^{m+2}}\right)J_{m+3/2}(\alpha r_1) \Bigr].
 \label{hrelchand}
 \end{eqnarray}
An equivalent formula for the relative helicity obtained using Equation (\ref{hrelb}) works out to be
\begin{equation}
H_{rel}(\alpha,n,m,r_1,r_2)=\frac{2}{\alpha}\int_V B_\phi^2 dV=\frac{8\pi\alpha(m+1)(m+2)}{2m+3}\int_{r_1}^{r_2}rJ_{m+3/2}^2(\alpha r)dr.
\label{hrelcb}
\end{equation}
The analytical equivalence of Equations (\ref{hrelchand}) and (\ref{hrelcb}) is presented in Appendix \ref{fabc}.

\section{\uppercase{The $\lowercase{r^2f^{2\beta-2}=1}$ case: LL modes }}
\label{llsol}
We now study the second set of solutions obtained in Section \ref{modes}: the LL modes.
The condition $r^2f^{2\beta-2}=1$ along with Equation (\ref{fpsep}) imply
\begin{equation}
 f^{\beta-1}=r^{-1}.\label{fb}
\end{equation}
Assuming the functional form
\begin{equation}
f(r)=r^{-n},\label{fm}
\end{equation}
where $n$ is a constant, gives the condition $\beta=(n+1)/n$, and Equation
(\ref{fpsep}) then yields the following equation as obtained by \citet{low90}:
\begin{eqnarray}
 r^2\frac{f^{\prime\prime}}{f}&=&n(n+1)\\
(1-\mu^2)P^{\prime\prime}+a^2\frac{n+1}{n}P^{1+2/n}+n(n+1)P&=&0.
\label{peq}
\end{eqnarray}
There is an arbitrary amplitude of $P$ in Equation (\ref{peq}) that can be scaled away.
Then this ordinary differential Equation (ODE) together with the homogeneous boundary conditions pose an eigenvalue problem
to determine the scaled parameter $a$ as an eigenvalue.
Recalling  Equation (\ref{spb}), we have the following expression for the nonlinear force-free modes:
\begin{equation}
 \mathbf{B}= \frac{-1}{r \sqrt{1-\mu^2}}\left(\frac{\sqrt{1-\mu^2}}{r}\frac{\partial \psi}{\partial\mu}\hat{\mathbf{r}}+
\frac{\partial \psi}{\partial r}\hat{\boldsymbol{\theta}}-Q\hat{\boldsymbol{\phi}}\right)
\label{low}
\end{equation}
where $\psi=P(\mu)/r^n$ and $Q=a\psi^{(n+1)/n}$. Now $P$ is the solution of Equations (\ref{peq}) and
(\ref{low}) and takes the form
\begin{equation}
 \mathbf{B}= \left(\frac{-1}{r^{n+2}}\frac{dP}{\partial\mu},
\frac{n}{r^{n+2}}\frac{P}{(1-\mu^2)^{1/2}},\frac{a}{r^{n+2}}\frac{P^{(n+1)/n}}{(1-\mu^2)^{1/2}}\right).
\label{Blow}
\end{equation}
Eqn (\ref{peq}) is not straightforward  to integrate numerically except for the case $n=1$, 
which was presented in \citet{low90}, because there is an inherent singularity at $\mu=0$. 
We extend these solutions to higher values of odd $n$ by using the following transformation:
\begin{equation}
 P(\mu)= (1- \mu^2)^{1/2} F(\mu),
\end{equation}
through which Equation (\ref{Blow}) stands as
\begin{equation}
\mathbf{B}= \left(\frac{-1}{r^{n+2}}\left[(1-\mu^2)^{1/2}F^\prime(\mu)-\frac{\mu F(\mu)}{(1-\mu^2)^{1/2}}\right],
\frac{n}{r^{n+2}}F,\frac{a}{r^{n+2}}(1-\mu^2)^{1/2n}F^{1+1/n}\right)
\label{flow}
 \end{equation}
and Equation (\ref{peq}) becomes
\begin{equation}
 (1-\mu^2)F^{\prime\prime}(\mu)-2\mu F^\prime(\mu)+\left[n(n+1)-\frac{1}{(1-\mu^2)}\right]
F(\mu)+a^2\frac{(n+1)}{n}F^{\frac{(n+2)}{n}}(1-\mu^2)^\frac{1}{n}=0.
\label{feq}
\end{equation}

Eqn (\ref{feq}) can be solved for all values of positive $n$, which represent the physically interesting cases. 
The initial requirement of $\psi=P/r^n$ requires only $n>0$
where $n$ can take any positive rational or integral value. A stringent condition
on $n$ is enforced if we demand $Q=a \psi^{1+1/n}$ is real, which is required for physically acceptable solutions.
 This means that for rational values 
of $n=p/q$, $Q=a \psi\psi^{q/p}$ and $\psi^q$ should be positive for all even values of $p$.
 Also, this implies that possibilities like (odd $p$, 
odd $q$) and (even $q$, odd $p$) are permissible. In summary, solutions exist for all odd values 
of $p$, whereas for even $p$, it exists only if $F(\mu)>0$ in the domain $-1 \leq \mu \leq 1$.
As examples, the solutions for $n=2/3, 2/5, 4/7$, etc., are allowed. Therefore the acceptable values of $n$ form a 
sufficiently dense set in the range $0 < n <\infty$, such that one can find instances of $p/q$ 
arbitrarily close to a given $n$. Recently, semi analytic solutions to Equation (\ref{peq}) for
 $n= 5,7...~201$ under the assumption of self-similarity were presented in \citet{zhang12}.

\subsection{Conditions for Physically Acceptable Solutions}
\label{llconds}
\begin{figure}[h!]
\centerline{\includegraphics[scale=.35]{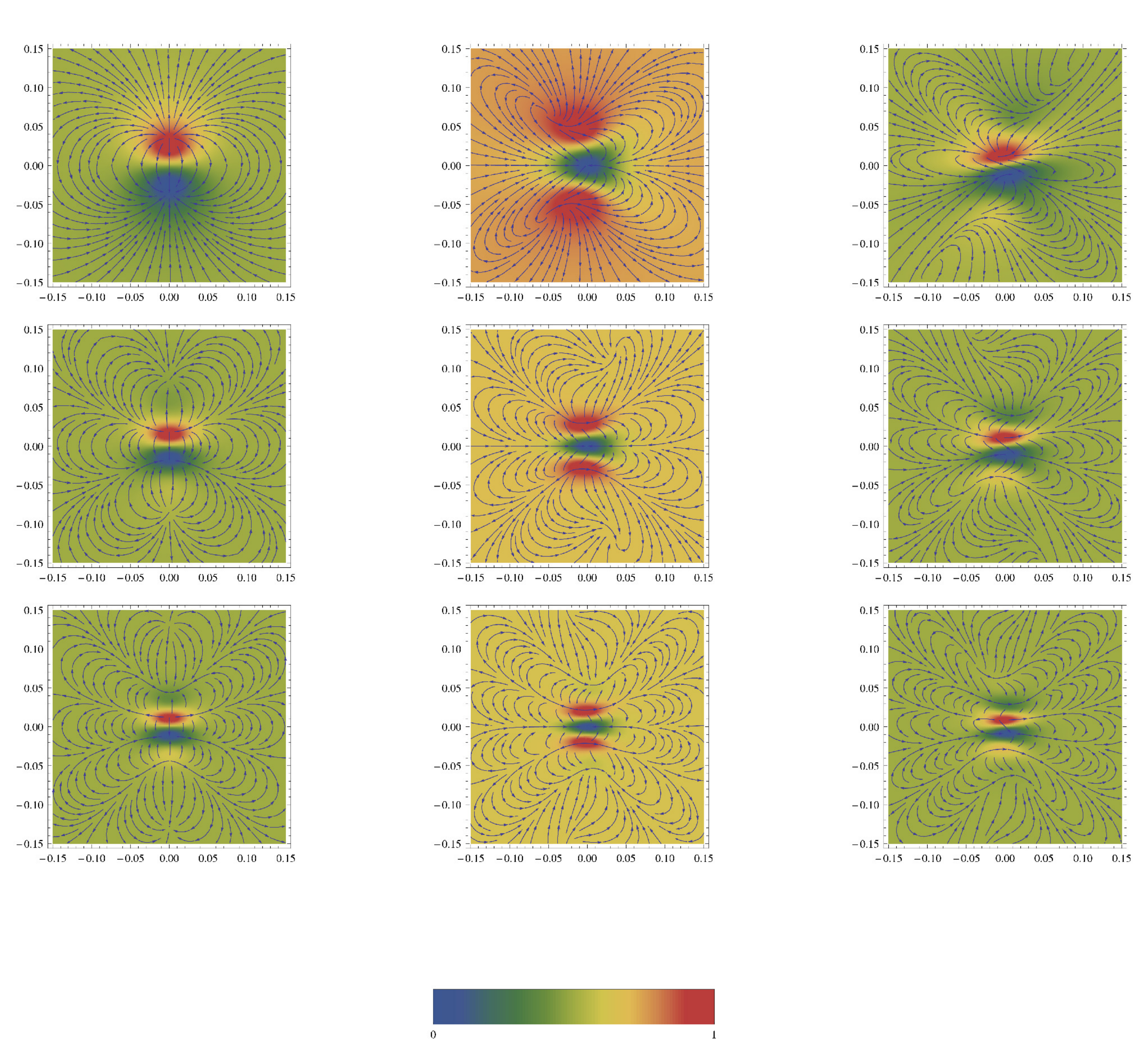}}
\caption{Sections (avoiding the origin where the fields are singular) shown are taken perpendicular
 to the radius at $r=0.05$ for different solutions of LL, with angular modes m = 1 
to $m = 3$ (columns) and radial modes $n=1$ to $n=3$ (rows).  
The contours represent the magnetic field lines projected on the plane
and the density plot represents the strength of the field perpendicular to the
plane of the figure. The values 0 and 1 in the color bar correspond to the minimum and maximum 
values of the perpendicular magnetic field, respectively.}
\label{llfig1}
  \end{figure}

The following conditions are required to be satisfied to obtain physically acceptable solutions:
\begin{enumerate}
 \item The field should be finite as $r\rightarrow\infty$. This is ensured if $n$ is positive.
\item The field should be well behaved and finite along the axis of symmetry. Because we have
\begin{displaymath}
 B_\theta, B_\phi\propto\frac{1}{(1-\mu^2)^{1/2}}P(\mu);
\end{displaymath}
 this requires that $P(\mu)$ should vanish at $\mu=1,-1$. This gives the condition
\begin{displaymath}
 P(\mu) \rightarrow (1-\mu^2)^s \quad \textrm{at}\quad \mu= \pm 1
\end{displaymath}
where $s \geq 1/2$.
\end{enumerate}
 The function $F(\mu)$ satisfies the boundary condition (see Appendix \ref{fbcond})
\begin{equation}
 F(\mu)=0\quad \textrm{at}\quad \mu= \pm 1.
\label{fbc}
\end{equation}
Further, it follows that Equation (\ref{feq}) reduces to the equation for the associated Legendre polynomials (where 
the last term is ignorable compared to the third term) in the limit $\mu^2\rightarrow 1$. 
So we can construct LL solutions by direct integration of Equation (\ref{feq}) subject to Equation (\ref{fbc}) for 
any positive $n$. We have cross-verified with the only case,  $n=1$, that was  given in \citet{low90} and calculated other allowed values of $n$ as well. 
The cases for the modes $n=1$ to $n=3$ are shown in
Figure \ref{llfig1} for the first three eigenvalues $m$ of the variable $a$. These
 solutions are singular at the origin, so the energy and helicity calculations are 
done excluding a spherical region around the origin.
As specific examples of the non integer type,
we depict realizations of radial modes  $n=7/5, 3/2, 9/5$, in the left column of Figure \ref{llfig2}.

\subsection{Energy and Relative Helicity for the LL Modes}
\begin{figure}[h!]
\centerline{\includegraphics[scale=.25]{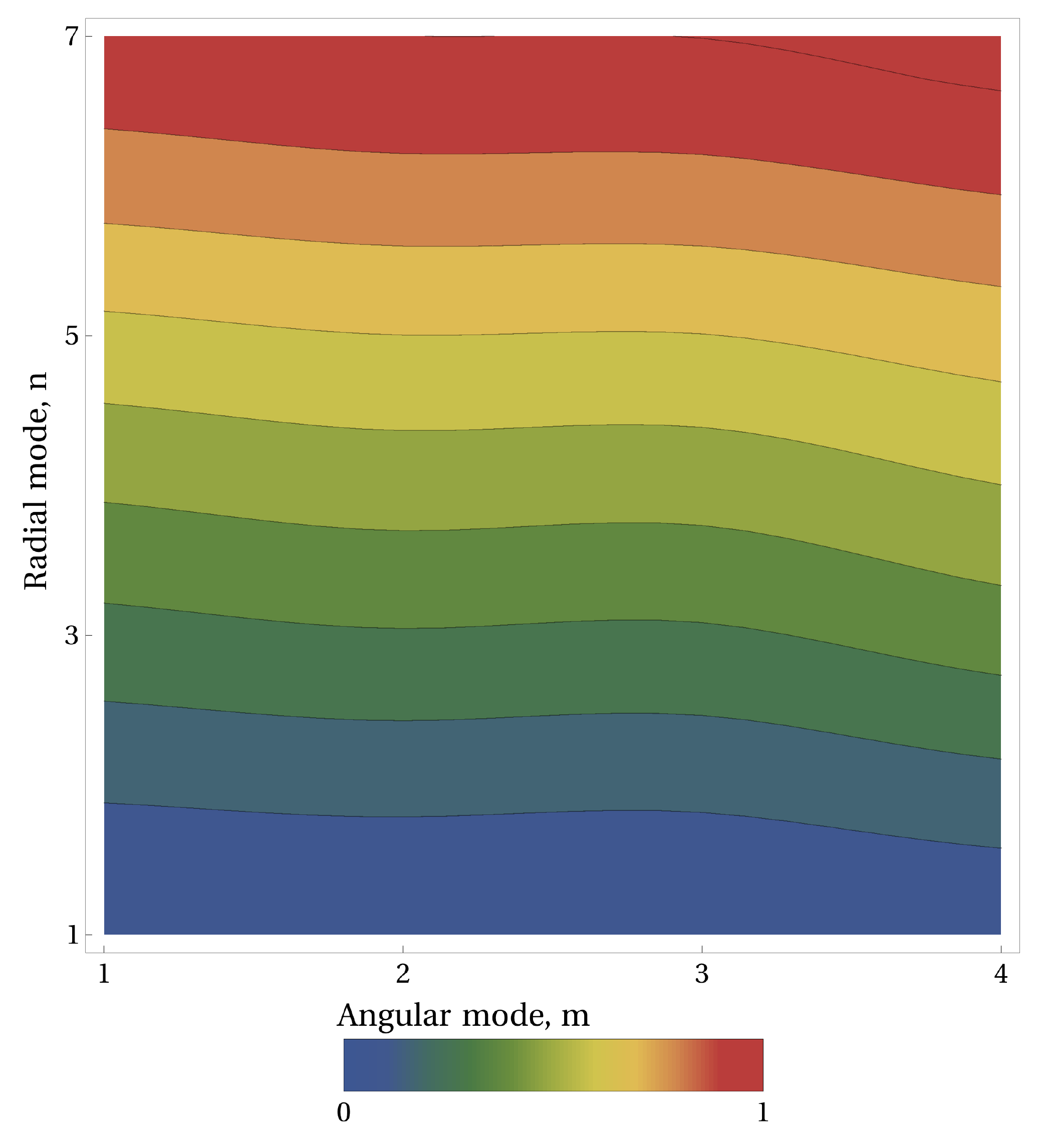}\includegraphics[scale=.45]{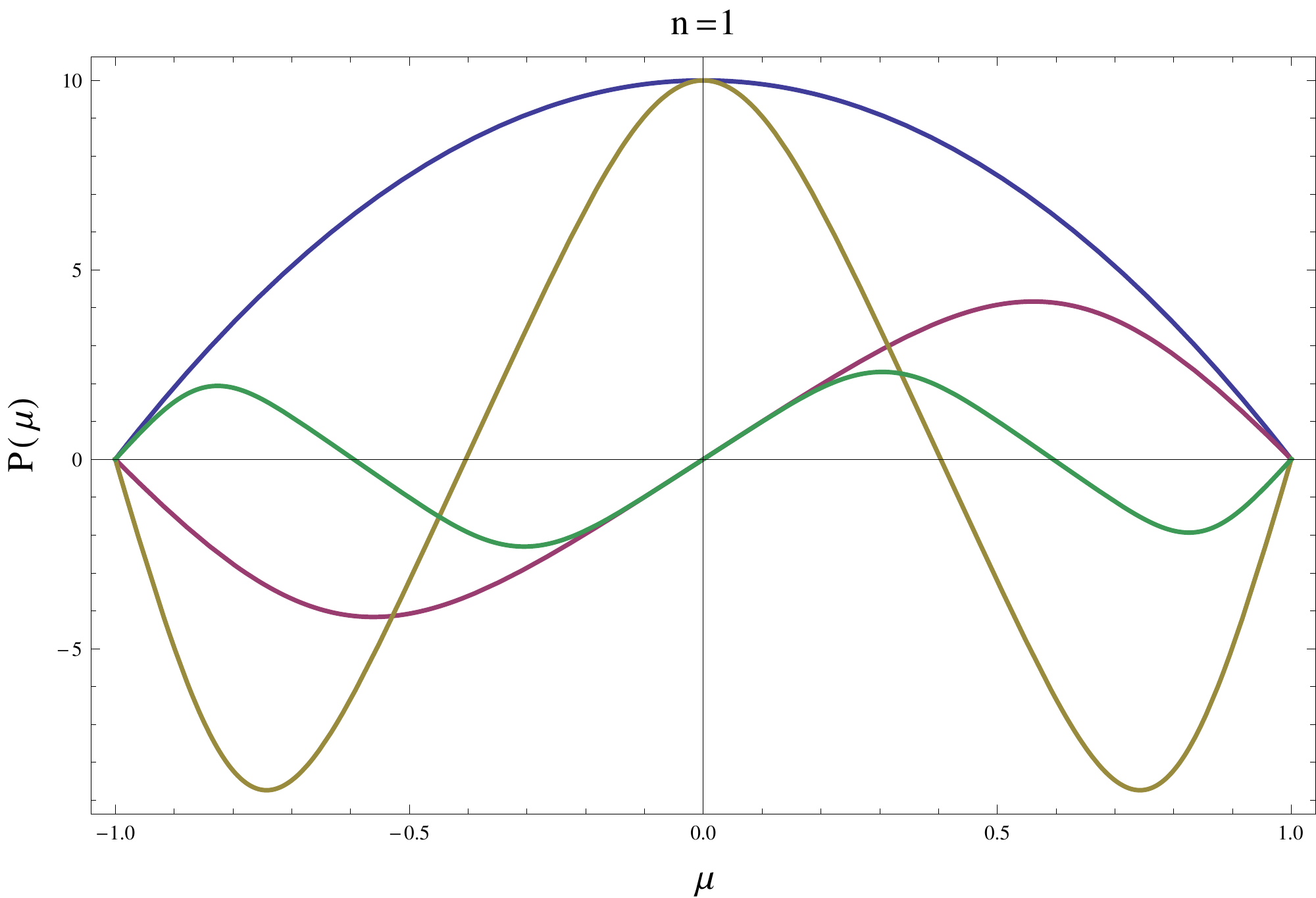}}
\centerline{\includegraphics[scale=.35]{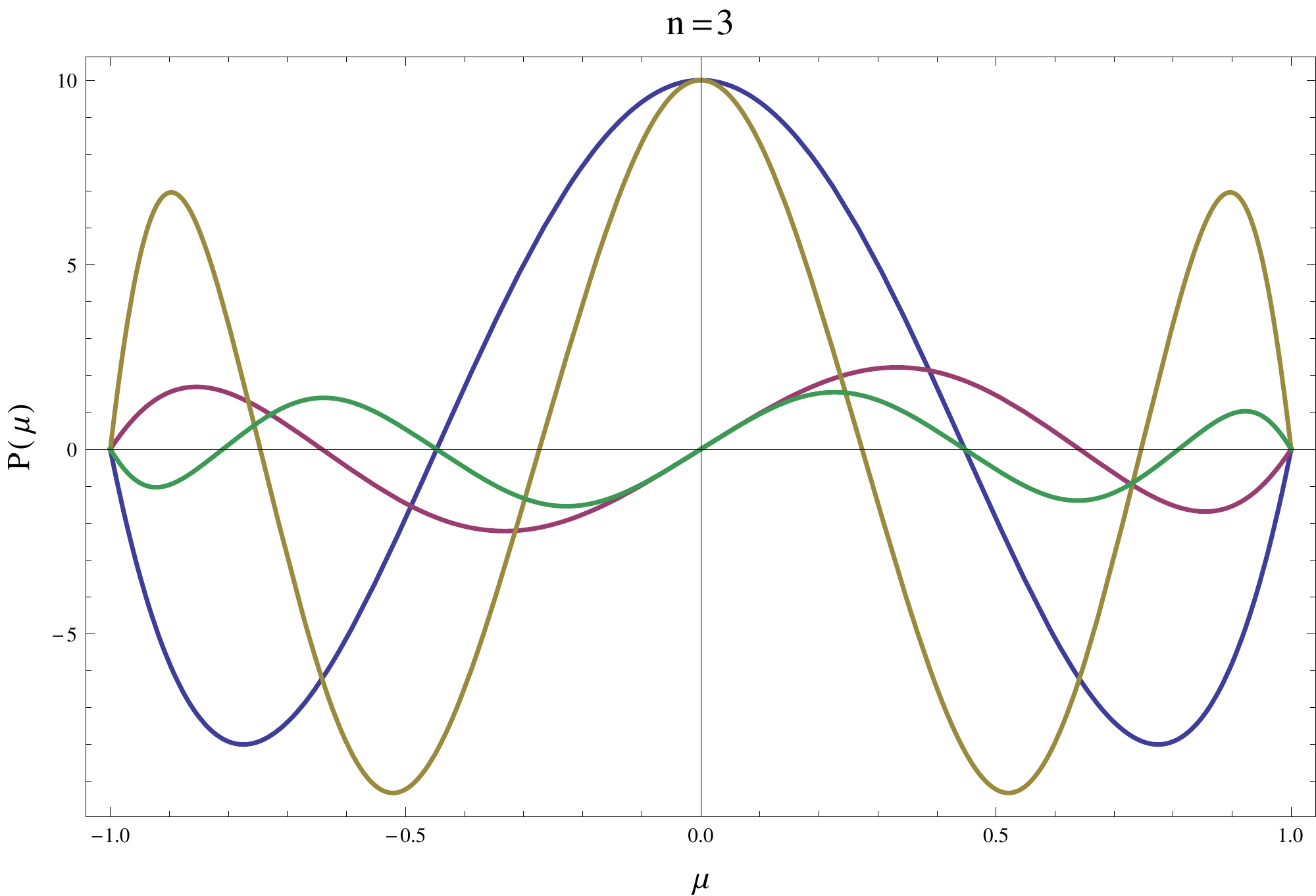}\includegraphics[scale=.35]{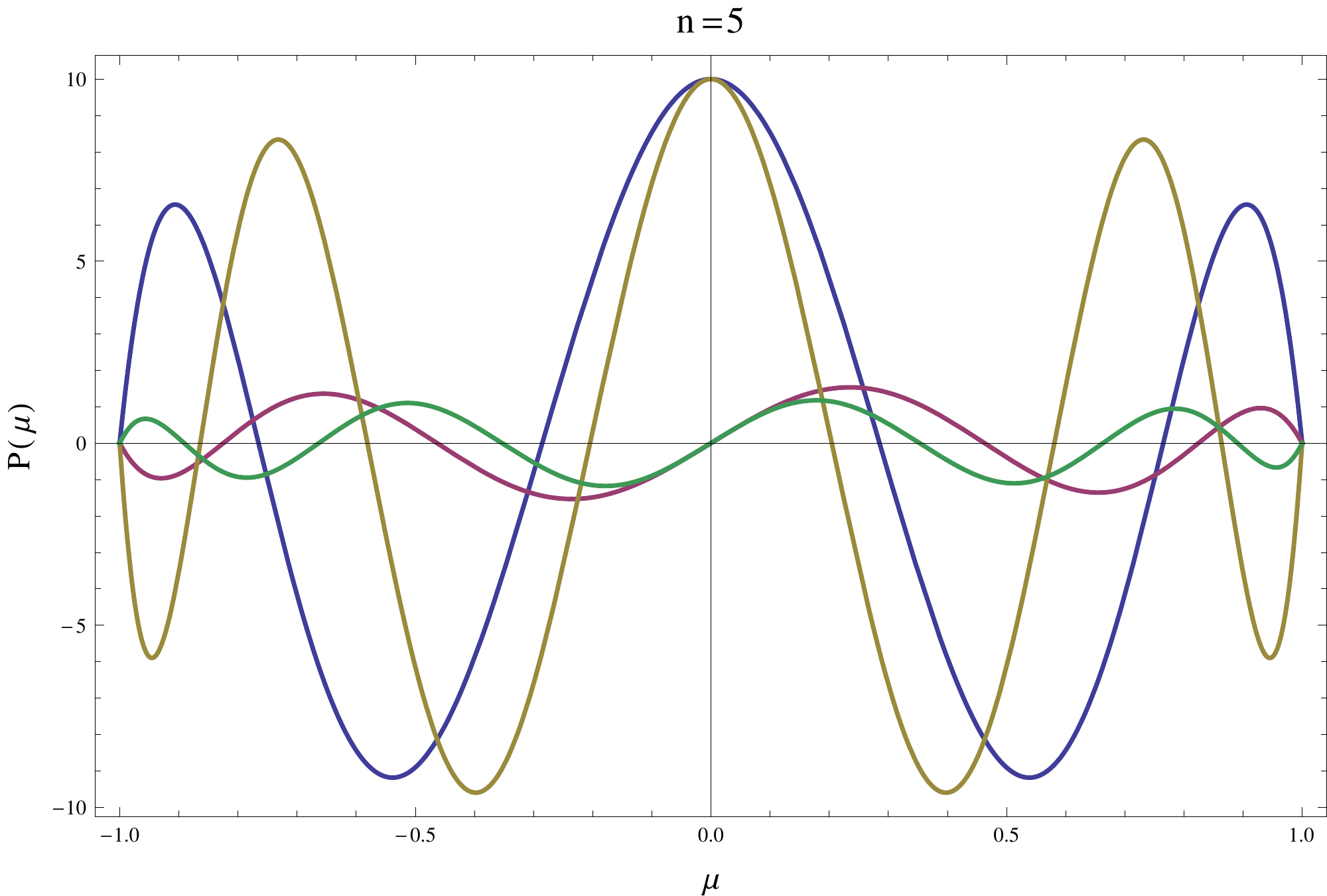}}
\caption{ Top left panel shows contours of energy for different angular and radial 
modes for the LL modes. The energies (normalized with respect to the maximum) 
are shown for different modes at the same lower boundary $r_1=0.5$. 
The value of contours are given in logarithmic scale because the parametric 
dependence is very sharp. The values 0 and 1 in the color bar correspond to the minimum and maximum 
values of energy, respectively. The next three panels show the realizations of $P(\mu)$
for the cases of $m=1$ to 4 for $n=1,3$ and 5. Note that the number of 
polarities for a given $(n,m)$ set is given by
$n+m-1$.}
\label{ell}
\end{figure}

The energy in the magnetic field is given by Equation (\ref{eff1})
\begin{equation}
 E_{\mathrm{ff}}(n,m, r_1)=\frac{1}{4(2n+1)r_1^{2n+1}}\int_{-1}^1d\mu
\left[ P^\prime(\mu)^2+\frac{n^2 P(\mu)^2}{1-\mu^2}+\frac{a^2 P(\mu)^{(2n+2)/n}}{1-\mu^2}\right]
\label{efflow}
\end{equation}
where the expression for the field from Equation (\ref{Blow}) is used. 
The energy of the force-free field calculated using the virial theorem of Equation (\ref{eviri}) gives the equivalent 
expression
\begin{equation}
 E_{\mathrm{ff}}=\frac{1}{4 r_1^{2n+1}}\int_{-1}^1\left\{\left(\frac{d P}{d \mu} \right )^2
-\frac{(n^2+a^2 P^{2/n})P^2}{(1-\mu^2)}\right\}d \mu.
\label{effvll}
\end{equation}
The Equation (\ref{efflow}) reduces to Equation (\ref{effvll}) by the use of Equation (\ref{peq}).
It may be noted that the function $P(\mu)$ is implicitly
dependent on the parameters $n$ and $m$.
The contour plot in Figure \ref{ell} shows the dependence of energy on the variables $n$ and $m$
and we find that the magnetic energy of the field increases with both of the
variables. The change in energy is very sharp with $n$ as compared to $m$, so the 
value of contours  are given in logarithmic scale.
The potential field corresponding to the LL mode is calculated in Appendix \ref{llpot},
and its final expression is given by
\begin{equation}
 \mathbf{B}_P=\left(\sum_{l=0}^{\infty}-(l+1)\frac{b_l}{r^{l+2}}P_l(\mu),\sum_{l=0}^{\infty}\frac{-b_l}{r^{l+2}}
(1-\mu^2)^{1/2}\frac{dP_l}{d\mu},0\right),
\label{Bplow1}
\end{equation}
where 
\begin{equation} 
a_l=0,\quad b_l=\frac{2l+1}{2(l+1)}r_1^{l-n}\int_{-1}^1\frac{dP}{d\mu}P_l(\mu)d\mu.
\label{alLL1}
\end{equation}
The energy for the potential field constructed from the LL modes (see Section \ref{llpot}) is given by
\begin{equation}
 E_{pot}(l,r_1)=\sum_{l=0}^\infty\frac{b_l^2(l+1)}{2(2l+1)r_1^{2l+1}}.
\label{epotlow}
\end{equation}
In order to calculate the relative helicity, we find the vector potential for the LL modes in Appendix \ref{llvpot}
given by
\begin{equation}
 \mathbf{A}=\left(0,\frac{-a}{n r^{n+1}}\frac{P(\mu)^{(n+1)/n}}{(1-\mu^2)^{1/2}},\frac{1}{r^{n+1}}\frac{P(\mu)}{(1-\mu^2)^{1/2}}\right).
\label{alow1}
\end{equation}
The vector potential for the potential field is given by Equation (\ref{Apchand}) with $a_l$ and $b_l$ as defined in
Equation (\ref{alLL1}).
Then the relative helicity is calculated from Equation (\ref{hrel}) to be
\begin{equation}
 H_{\mathrm{rel}}(n,m,r_1)=\int_V(\mathbf{A}_P\cdot\mathbf{B}- \mathbf{A}\cdot\mathbf{B}_P)dV
\label{llhr1}
\end{equation}
because $\mathbf{A}\cdot\mathbf{B}=\mathbf{A}_P\cdot\mathbf{B}_P=0$. 
Thus, even if the absolute helicity $\mathbf{A}\cdot\mathbf{B}$ is zero in our
chosen gauge, the cross terms in the Finn\textendash Antonsen formula give rise to
 the nonzero values of the relative helicity.
The expression in Equation (\ref{llhr1}) can be further simplified to
\begin{equation}
 H_{\mathrm{rel}}(n,m,r_1)=-2\pi a \sum_{l=0}^\infty \int_{-1}^1  \frac{b_l}{n l r_1^{n+l}}P^{1+1/n} \frac{d P_l}
{d \mu}d\mu.
\label{hrellow}
\end{equation}
Using Equation (\ref{hrelb}), we get an equivalent form for the relative helicity
 that works out to be
\begin{equation}
 H_{\mathrm{rel}}(n,m,r_1)=\frac{2\pi a}{n r_1^{2n}} \int_{-1}^1  \frac {P^{2+1/n}}{(1-\mu^2)}d\mu. 
\label{hrbll}
\end{equation}
The two formulae in Equations (\ref{hrellow}) and (\ref{hrbll}) are equivalent as shown 
in Appendix \ref{fabll}.

\section{\uppercase{Simulation of magnetograms}}
\label{s:simulation}

\begin{figure}[ht!]
\centerline{\includegraphics[scale=0.4]{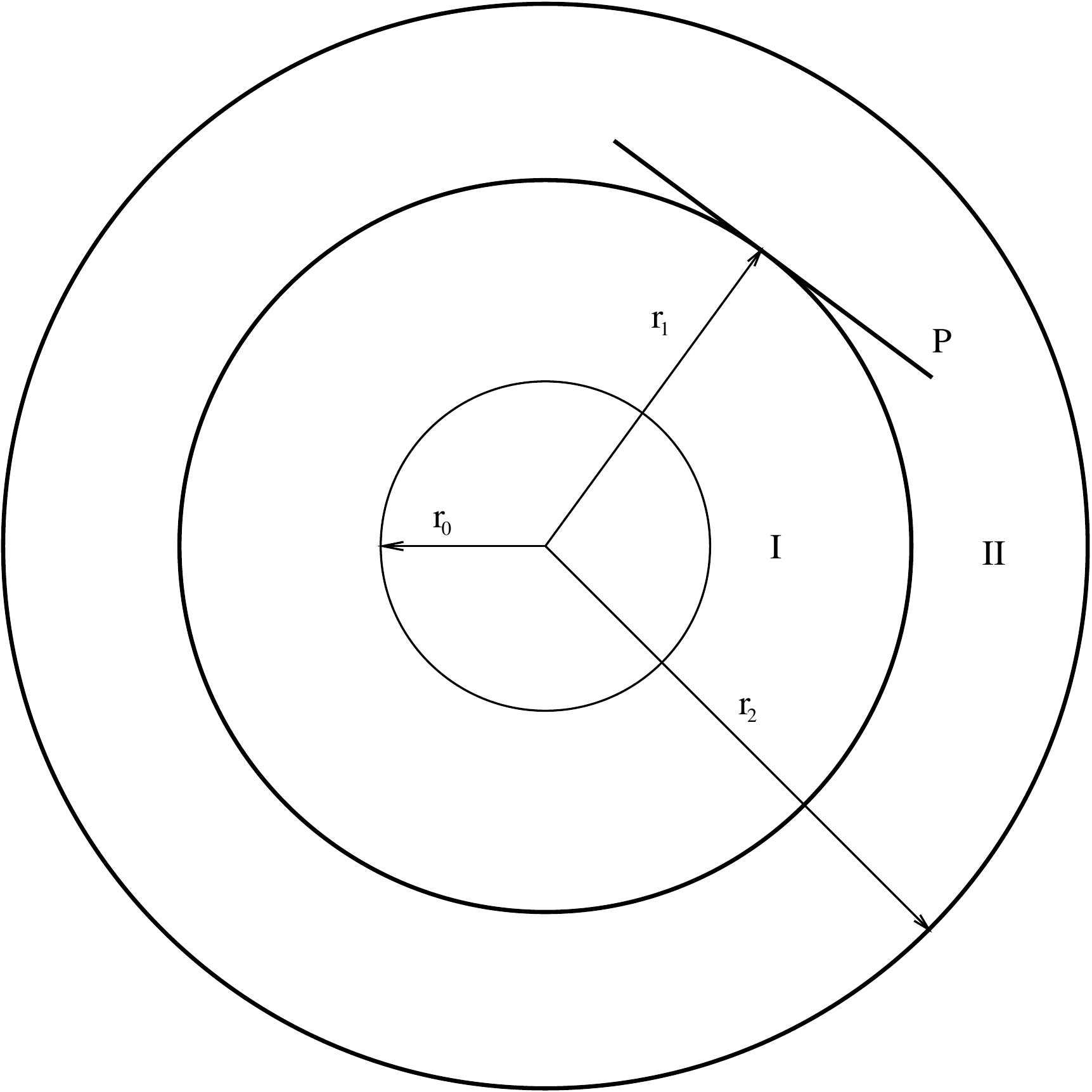}\quad \includegraphics[scale=0.55]{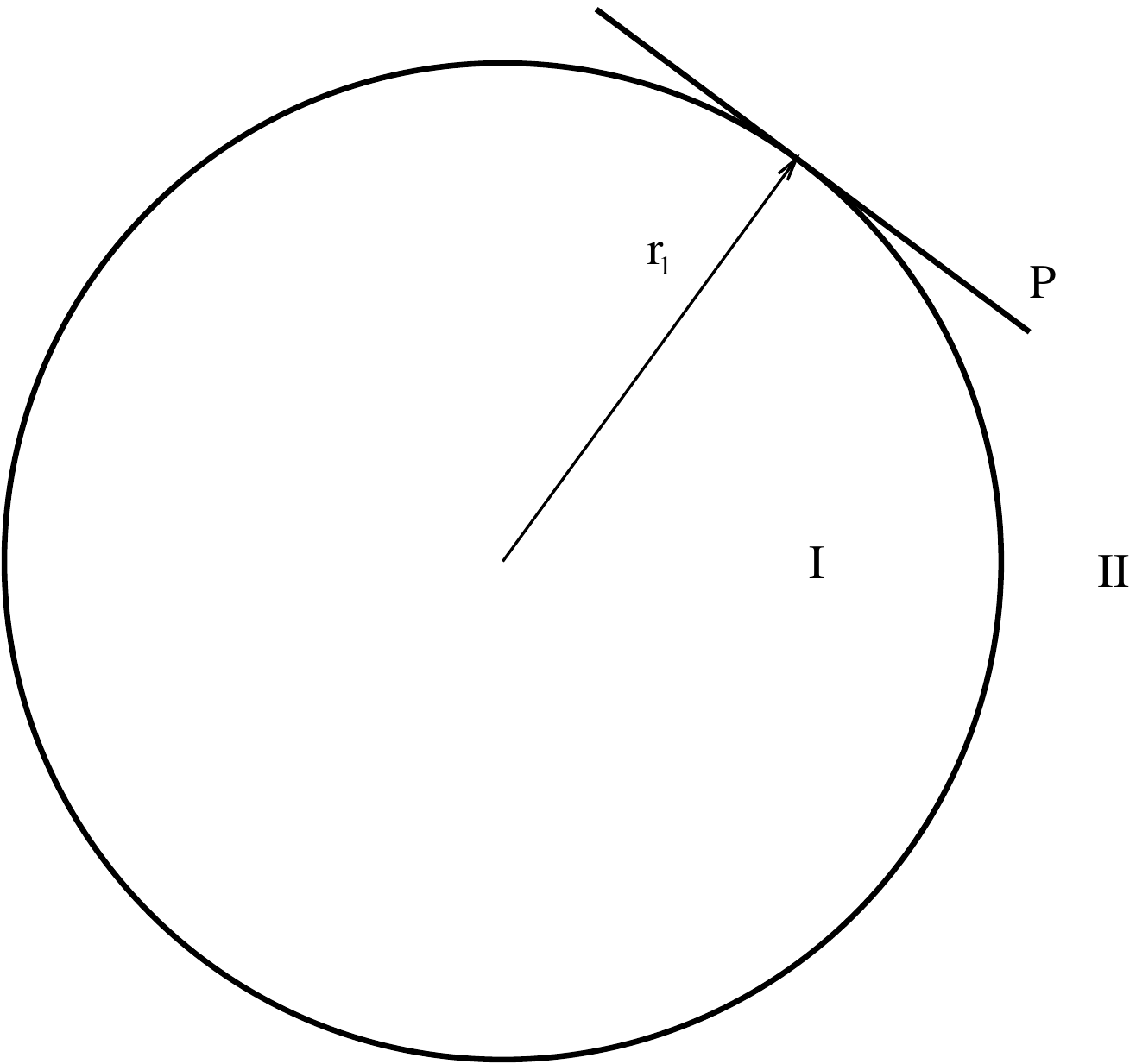}}
\caption{ Geometry used in the problem for the linear (left panel) and
nonlinear (right panel) fields. For the linear case, the field is first computed between 
radii $r_0$ and $r_2$. A plane representing the magnetogram is placed tangential to
a shell of radius $r_1$. A potential field is constructed in the spherical shell between
 radii $r_1$ and $r_2$ (region $II$) using the normal components of the force-free field
 at the lower boundary, $r_1$. For the nonlinear case, the field is computed
outside a shell of radius $r_1$ where the plane representing the magnetogram is placed tangentially.
 Again the potential field is constructed using the normal components of the force-free field
 at $r_1$.}
\label{bound}
\end{figure}

In this paper, our aim is to get reasonably good and quick estimates of free energy and relative
helicity for the active region observed in the magnetograms.
It is well known that NLFF fields best represent the solar active regions, and the most
useful and widely used analytic solution is the Low\textendash Lou solution in the spherical
geometry. Hence this geometry was naturally chosen. In our scheme, we first compute a
large set of linear and nonlinear 3D force-free modes in a spherical
shell volume where the magnetogram is a tangent plane to the lower boundary, see Figure \ref{bound}.

For the linear case, the field is defined between 
radii $r_0$ and $r_2$. A plane representing the magnetogram is placed tangential to
a shell of radius $r_1$. A potential field is constructed in the spherical shell between
 radii $r_1$ and $r_2$ (region II) using the normal components of the force-free field
 at the lower boundary, $r_1$. For the nonlinear case, the field is defined
outside a shell of radius $r_1$ (region II) where the plane representing the 
magnetogram is placed tangentially. Again the potential field is constructed using
 the normal components of the force-free field at $r_1$.
The fit to the magnetogram data selects a particular mode 
of force-free field in spherical geometry (details given below). Apart from this, the magnetogram also sets
a length scale  for the problem and fixes an amplitude of the magnetic field. 
Both the force-free field and potential field are known completely in region II,
so we calculate the free energy and relative helicity (using Finn\textendash Antonesen and Berger formula) in region II.
In order to compare with the other estimates available in the literature, where the potential fields
are usually extended from the planar surface of the magnetogram to a cuboidal
volume over the magnetogram, we rescale our physical quantities obtained for a hemisphere
by the factor of the solid angle subtended by the magnetogram at the center. 
We would like to emphasize that our problem is to reconstruct the entire field from the knowledge of
the field in a 2D plane, which does not a priori force any choice of geometry.
So calculation of free energy and relative helicity in the shell geometry
does not compromise our original goal, which is to get quick reasonable estimates of these quantities
over the solid angle subtended by the magnetogram.
The validity of this approximation can be seen from the general agreement with 
other estimates (including observations, presented later in Table \ref{compare}). An advantage
in this method is its ease and utility in calculating these physical quantities: in particular
the relative helicity is thus far not calculated by other approaches. Further,
 we do not have to assume any other boundary condition for the side walls, as  required
in the other extrapolation techniques using the cuboidal volume.

\begin{figure}[h!]
\centerline{\includegraphics[scale=.35]{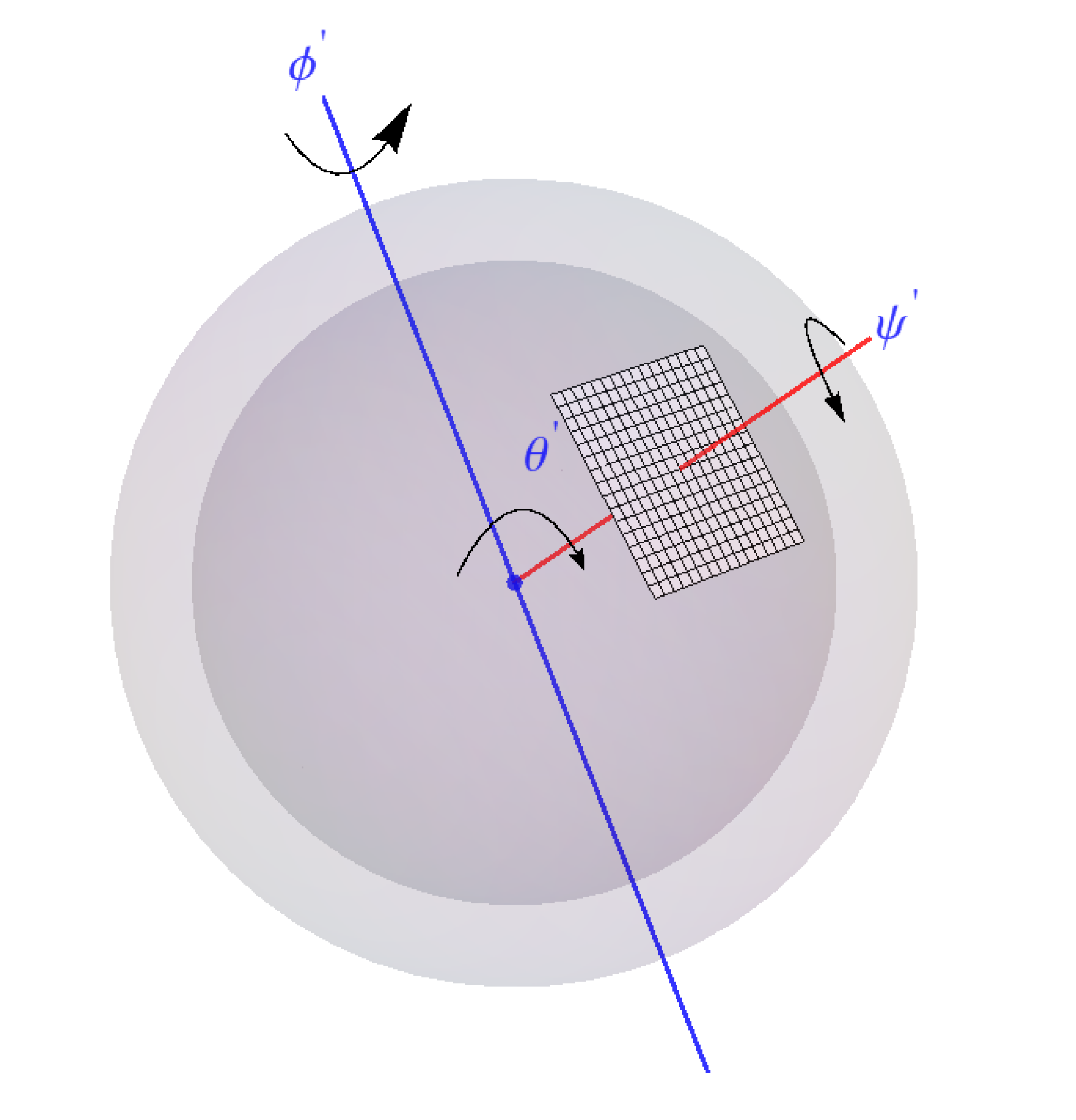}}
\caption{Magnetogram is simulated by taking a cross section of the axisymmetric 3D force-free field
at a radius $r_1$.The magnetogram is then rotated through the Euler angles $\theta^\prime$ and $\psi^\prime$
to match the components of the observed magnetogram. The rotation $\phi^\prime$ is redundant
because the field is axisymmetric.}
\label{shellfig}
\end{figure}

We now use the library of LL and C modes by taking 2D sections of the 
force-free spheres appropriately and compare these sections with the
observed magnetograms. We describe the best-fit mode and figure of merit of fit in Section \ref{fit}. 
The following steps are taken in simulating the sections:

\begin{enumerate}
\item We compute the 3D force-free magnetic field in spherical geometry corresponding to a given C and  LL mode from
Equation (\ref{bchand}) and Equation (\ref{Blow}) respectively. 

\item
The coordinates on the magnetogram are labeled as the $x$ and $y$ axes, where the $x,y  \in [-0.5,0.5]$
so that the magnetogram is of unit length.

\item A cross section of the sphere is taken at a radius $r_1$, and all three components of magnetic field are
computed over this 2D surface, see Figure \ref{shellfig}. The orientation of the magnetogram is given by the
three Euler angles $(\phi', \theta', \psi')$, of which the angle $\phi'$ is redundant because the fields are
axisymmetric. The transformation matrix for the Euler rotation is given by
\begin{equation}
 \Lambda(\theta^\prime,\psi^\prime)=\begin{bmatrix}\cos\psi^\prime  &\cos\theta^\prime\sin\psi^\prime  &\sin\psi^\prime\sin\theta^\prime  \\
 -\sin\psi^\prime  &\cos\theta^\prime\cos\psi^\prime   &\cos\psi^\prime\sin\theta^\prime \\ 0 &-\sin\theta^\prime   & \cos\theta^\prime \end{bmatrix}.
\end{equation}
 In effect the position and orientation of the section is fixed by three parameters $(r_1, \theta', \psi')$.
We then transform a point on the magnetogram with coordinates $(x,y,z)$ by the
inverse of $\Lambda$. 
\item
The coordinates in spherical 
 $  \mathbf{x}_S \equiv (r,\theta,\phi)$ are obtained from Cartesian
coordinates $ \mathbf{x}_C \equiv (x,y,z)$ through the operator $S$ given by
\bea
\mathbf{x}_S&=& S (\mathbf{x}_C) \nonumber \\
r&=&(x^2+y^2+z^2)^{1/2} \nonumber\\
\phi &=&  \left \{ \begin{array}{ll} \arctan{(y/x)} & x >0 \\ \arctan{(y/x)} +\pi & x \leq 0 \end{array} \right.  \\ 
\theta&=&\arccos{ \left ( z/(x^2+y^2+z^2)^{1/2} \right )} \nonumber
\label{ctos}
\eea
 to get $\mathbf{x}_S$ as a function of $(x,y,z)$ so that we have the coordinates of all the points on the 
magnetogram in spherical coordinates
\be
\mathbf{x}_S=  S \left( \Lambda^{-1}(\theta', \psi')\mathbf{x}_C \right).
\ee

\item
We now evaluate the magnetic field in spherical coordinates 
$\mathbf{B}_S(\mathbf{x}_S)$ and then convert the components of
magnetic field from spherical polar to Cartesian coordinate system so that

\be
\mathbf{B}_C [\mathbf{x}_C,\theta', \psi', x, y]= \Lambda(\theta', \psi') T \left( \mathbf{B}_S \left[ S(
\Lambda^{-1}(\theta', \psi') \mathbf{x}_C) \right] \right).
\ee
where
\begin{equation}
 T=\begin{bmatrix} \sin\theta\cos\phi & \cos\theta\cos\phi & -\sin\phi \\ \sin\theta\sin\phi  & \cos\theta\sin\phi  &\cos\phi \\
\cos\theta & -\sin\theta & 0 \end{bmatrix},
\end{equation}
is the transformation from spherical to Cartesian.
Here the coordinates $\theta$ and $\phi$ are locations on the magnetogram computed 
from Equation (\ref{ctos}). Because $\Lambda$ and $T$ are orthogonal, 
their inverses are the corresponding transposes.

\end{enumerate}

\begin{figure}[h!]
 \centerline{\includegraphics[scale=.19]{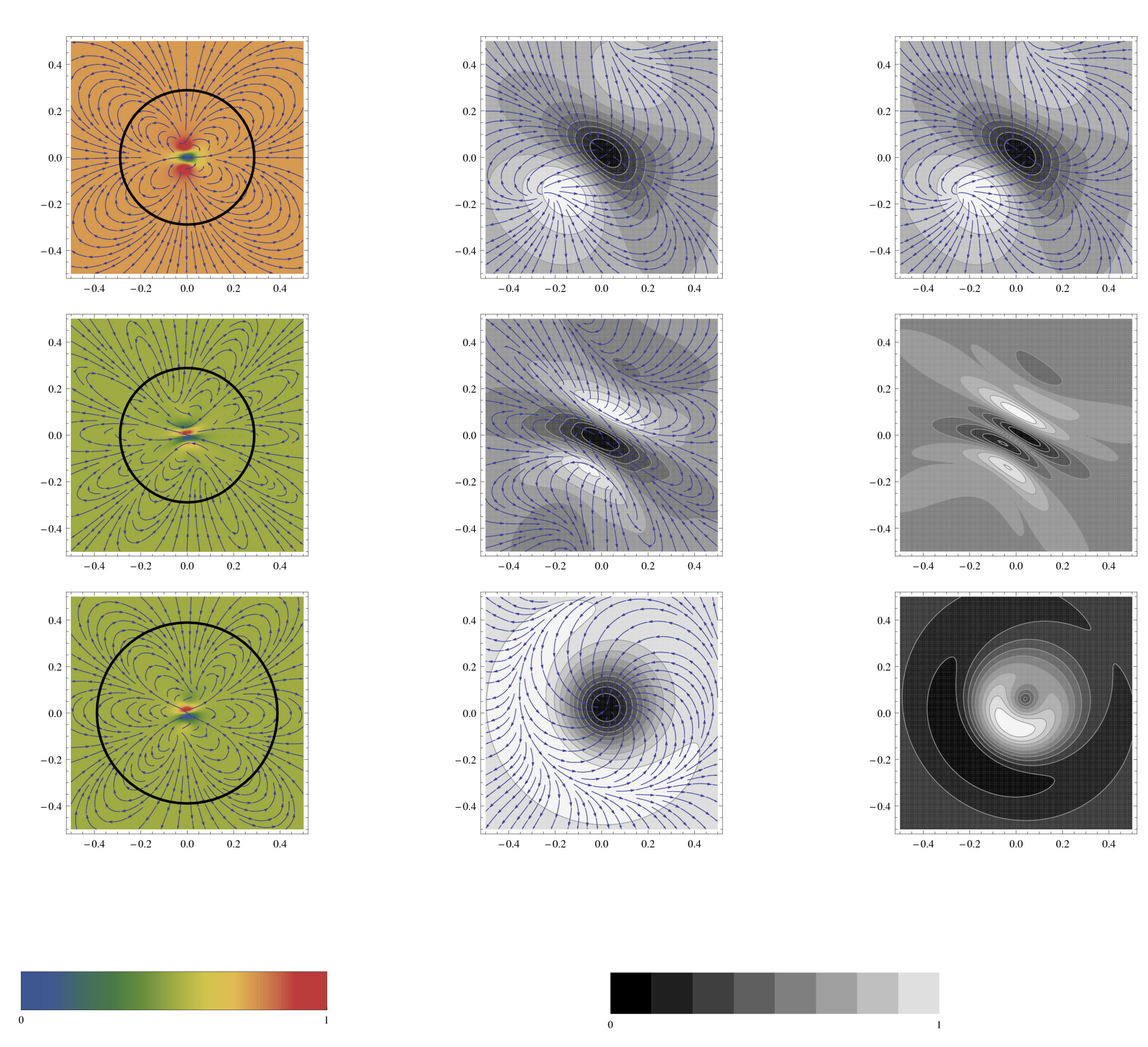}}
 \caption{Sections for different LL solutions, taken perpendicular to the radius at $r=0.05$ with  parameters
$(n, m, r_1, \theta, \phi)$=$\{$(7/5, 2, 0.29,1.75,4.14), (3/2, 3, 0.29,1.75,4.14), (9/5, 1, 0.39, 0.18, 4.14)$\}$ 
are shown in the top, middle and bottom rows respectively. In the left column, the contours represent
the magnetic field projected onto the plane of the figure,
and the density plot represents the strength of the field perpendicular to it.
The circles are drawn at radius $r_1$. The middle column is a section of the field and the right column is a section
of the resulting current density; for an illustration of the section geometry and the parameters,
 see Figure \ref{shellfig}.  The legends representing the strength of component normal to the page are shown below 
corresponding to the left and middle/right columns. The 0 and 1 in the legend scale correspond to the minimum and maximum 
values of the normal magnetic field respectively.}
\label{llfig2}
  \end{figure}

We illustrate simulated sections thus generated for the LL modes $(n, m) =\{(7/5, 2)$, $(3/2, 3)$, $(9/5,1)\}$
in Figure \ref{llfig2}. The parameter values $(r_1, \theta, \phi)$ chosen are indicated 
in the caption and the resulting sections are typical of the single- and double-polarity active regions seen in observations.

\subsection{Search Strategy}
\label{search}
We have the following free parameters in the problem:
\begin{enumerate}
 \item The radial and angular mode numbers, $n$ and $m$: the values for $n$ and $m$ fix the force-free modes 
(for both C and LL solutions). Whereas $n$ takes only integral values for C modes, LL modes can take integral 
as well as fractional values (with the exceptions mentioned in $Section \ref{llsol}$). The value of $m$ takes 
only integers for both C and LL modes.

\item The first derivative $d= F'(\mu=-1)$: the value of the derivative of $F(\mu)$ at the boundary which is
used as a boundary condition for solving Equation (\ref{feq}), is a free parameter; this only scales the
solution by an arbitrary constant. In this paper we have used $d=10$ as a constant 
input for all calculations.

\item Euler angles, $\theta^\prime$ and $\psi^\prime$:
The C modes repeat in $\theta^\prime$ at an interval
of $\pi/(m+1)$ for a given $m$ and $\theta^\prime$ was taken to be the larger of
this value and the angle subtended by the magnetogram at the center to avoid 
redundancy; $\theta^\prime= \max{ (\pi/(m+1), \arctan{(L/r_1)})}$. For LL modes we search in the domain $\theta^\prime\in [0,\pi]$. 
For both cases, we search for $\psi$ in the range, $\psi^\prime\in [0,2\pi] $. If the  magnetogram has $n_p$ 
polarities in a range $\Delta \mu=1-\cos\gamma$ for a mode that has $m_p$ total polarities over the domain $[-1,1]$, 
then we estimate
\be
 \frac{n_p}{m_p}\simeq \frac{\Delta\mu}{2}=\frac{1-\cos\gamma}{2}=1/(1+ (r_1/L)^2 4)
\label{np}
\ee
where $\gamma$ is the angle subtended by the magnetogram at the center.
\item 
The radius $r_1$ at which the cross section is to be taken is in the range 0 to $r_2$. Here 
$r_2$ is the outer radius up to which 
the energies and helicity are calculated. In the case of C modes $r_2$ is finite and necessarily at 
the root of $J(m+3/2, \alpha r)$. We have restricted 
the calculations to only one radial oscillation (corresponding to only one visible closed loop along 
the line of sight), whereas for LL modes $r_2$ is infinite as the fields tend to zero only at infinity. 
The finite radial boundary $r_2$ is needed for the C modes and not the LL modes.
For a linear force-free field in the Taylor theory of relaxation in the unbounded atmosphere, 
the minimum energy state is a C mode provided the domain is finite \citep{low96}.
In the case of C modes, the constraint that the magnetogram is contained within the sphere of radius $r_2$ leads to
the condition
\be
r_1 \leq \sqrt{ r_2^2 -L^2/4}.
\ee
 In the case of LL modes, there is no obvious constraint on $r_1$.

\item 
The force-free parameter $\alpha$: for C modes, $\alpha$ is a constant and has to be given as an input.
We restrict $\alpha^{-1}$ to be of order unity in line with typical observed magnetograms where the field
reverses over this length.
\end{enumerate}

To summarize, the parameter space to search for C modes is 
${(n, m, r_1, r_2, \alpha, \theta^\prime, \phi^\prime)}$.
We start sweeping from the lowest combinations
of $(n,m)$ in increasing energy and searching for $r_1$ with the constraint on the range of
$\theta^\prime$ and $r_1$ given above and allowing only for one radial oscillation.

For LL modes, we have to search for the parameters ${(n, m, r_1, \theta^\prime, \phi^\prime)}$. Here, we start by sweeping 
from the lowest combinations of $(n,m)$
and looking for $r_1$ near unity to find the best-fit lowest energy modes within the allowed range
of $\theta^\prime$ and $\psi^\prime$. Due to the computational constraints involved we were only able to survey a subset of the parameter space. The run time for a combination of $(n,m, \alpha)$ for C modes 
and $(n,m)$ for LL modes is about 8 hr of parallel computation on three computers with second generation Intel i7 processors.  The search for the best-fit parameter
is done in the following manner. For a particular mode of the solution 
(specified by the values of $n$, $m$, and
$\alpha$, in the case of C modes), we choose six equispaced grid points in the $\theta^\prime$ and $\psi^\prime$ domain and
eight equispaced points for $r$. For the C modes the values for $r$ are chosen between two Bessel zeros, whereas for the
LL modes we start with an initial guess of $r=1$. Then for each combination of $(r,\theta^\prime, \psi^\prime)$ the
field is computed over a 380$\times$380 grid (for a typical magnetogram). Thus, each mode of the solution involves evaluating 
the field  for about 42 million combinations. All of the template grids thus generated are compared with the observed data; following this initial search, the best-fit set is selected
and a finer grid of $(r,\theta^\prime, \psi^\prime)$ defined about this set with four grid points each
 is searched to obtain the final parameter set. We plan to expand upon the search in the future when we are able to 
make our code, which is already parallelized, run on a faster cluster.

\subsection{Fitting Parameters}
\label{fit}
In the previous section we described how
we explore the parameter space and 
generate a large ensemble of magnetograms. In order to select the best-fit
with the observations, we define a figure of merit, $c$   
for the magnetic field $\mathbf{B}$ as
\begin{equation*}
  c=\frac{\langle(\mathbf{B}_T\cdot\mathbf{B}_O)| \mathbf{B}_O|\rangle}{\langle|\mathbf{B}_T|^3
\rangle^{1/3}\langle|\mathbf{B}_O|^3\rangle^{2/3}},
\label{fitting}
\end{equation*}
which is the normalized dot product between the observed, $\mathbf{B}_O$, and the theoretically simulated field, 
$\mathbf{B}_T$, weighted by the strength of the observed magnetic field so that $|c|$ would be unity 
for a perfect correlation. Here $\langle \rangle$ represents the mean computed over the entire grid.
We also calculate the following correlation parameters to estimate the goodness of the fit for the selected configuration
\begin{equation}
 d=\frac{\langle (\mathbf{B}_{O}\cdot\mathbf{B}_{T}/|\mathbf{B}_{T}|)\rangle}{\langle |\mathbf{B}_{O}| \rangle}, ~~~~~~~~{\rm and}
\end{equation}
\begin{equation}
\epsilon =\frac{\langle |\mathbf{B}_{T}|^2 \rangle}{\langle |\mathbf{B}_{O}|^2 \rangle},
\end{equation}
where $d$ is the average of the cosine of the angle between the two fields computed over the entire grid, which is
normalized by the strength of the observed field, whereas $\epsilon$ is the ratio of the magnetic energies
of the theoretical and observed fields. The amplitude of the theoretical field is set by multiplying a scaling constant, $g$, where
\begin{equation}
 g^2=\frac{\langle| \mathbf{B}_O|^3\rangle/\langle| \mathbf{B}_O|\rangle}
{\langle|\mathbf{B}_T|^3\rangle/\langle|\mathbf{B}_T|\rangle},
\end{equation}
 which is deduced from the weighted ratio of energies.
Because the energy and helicity are computed for the entire sphere, we need to scale down these
quantities by the fraction of solid angle subtended by the magnetogram. 
The fraction of solid angle subtended by a square loop of size $L$ placed at a distance $r_1$
from the center is given by
\begin{eqnarray}
 \Omega_f&=&\frac{1}{\pi}\int_0^{L/2}\int_0^{L/2}\frac{r_1~dx~dy }{(r_1^2+x^2+y^2)^{3/2}}=\frac{1}{\pi}\int_0^{L/2}\frac{dx}{r_1^2+x^2}\int_0^{L/(2\sqrt{x^2+r_1^2})}\frac{r_1~dz}{(1+z^2)^{3/2}} \nonumber\\
&=&\frac{L r_1}{2 \pi}\int_0^{L/2}\frac{dx}{(r_1^2+x^2)\sqrt{L^2/4 +r_1^2+x^2}}=\frac{1}{\pi}\int_0^{\arctan{[L/(2r_1)]}}\frac{\cos{\theta}d\theta}{\sqrt{\cos{\theta}^2 +4r_1^2/L^2}}\nonumber\\
&=&\frac{1}{\pi}\arcsin\left(\frac{L^2}{L^2+4r_1^2}\right).
\end{eqnarray}
The final expressions for energy and helicity are given by
\begin{equation}
 \overline{E}=E\Omega_f g^2 L^3 \quad \textrm{and}\quad \overline{H}_{rel}= H_{\mathrm{rel}}\Omega_f g^2 L^4, 
\end{equation}
where $\overline{E}$ and $\overline{H}_{rel}$ represent the energy and relative helicity,
respectively, calculated over the volume containing the magnetogram.
\subsection{Effectiveness of the Search Strategy}
\label{effective}
In order to estimate the effectiveness of our search strategy, 
we try to recover the field configurations and energies of known input fields. 
The input fields used as test cases are 
\begin{enumerate}
 \item a pure dipole field,
\item an axisymmetric linear force-free field (C modes), and
\item a non axisymmetric linear force-free field, \citet{kendall57} (CK modes).
\end{enumerate}

In each of these cases, we gave a 2D cross section of the magnetic field as an input to our code and
obtained a best-fit with axisymmetric NLFFs (LL modes).
The parameter search grid used for this analysis is the same as that used for the observed field.
The details of the fit and the comparison of energy and helicity are presented in Table \ref{testtab}. 

 \begin{table}[h!]
\label{data}
\resizebox{18cm}{1.3cm}{
\begin{tabular}{|c|c|c|c|c|c|c|c|}
 \hline
No.&  Test Field& Mode & Correlation, c & Energy  & Energy 
& Relative Helicity & Relative Helicity    \\
&  &$n$, $m$ &  &  (input field) &  (best-fit field
)& (input field) &  (best-fit field)   \\

 \hline
a. & Dipole field&1, 1& 0.9925 & 7.77 & 6.76 & 0 & 0   \\
 
b. & C mode &3, 1& 0.662 & 1.24 & 1.34 & -4.97 & 0  \\

c.& CK mode  & 3, 1& 0.636 & 0.237 & 0.134 & -0.327 & 0  \\
 
\hline
\end{tabular}
}
\caption{ The table presents the correlation parameter for fits of the input 
test field with the LL modes along with a comparison of energy and relative 
helicity. In all the cases above the length scale  of the 2D cross section
is taken to be unity.}
\label{testtab}
\end{table}

The summary of our investigations can be presented as follows.
\begin{enumerate}
 \item \textit{Pure dipole field}. We find that the dipolar field gives an almost exact fit to the LL mode.
This is because it is an exact solution to the $n=1$, $m=1$ LL mode. The accuracy of the fit can be 
improved by taking more grid points in our parameter space. In this case, the axis of symmetry does not match exactly 
due to the smaller grid resolution that is chosen because of numerical constraints. Of course when the exact values 
of the parameters are chosen, we recover perfect fits. We also see that the energy of the
best-fit field closely matches that of the original field. 
\item \textit{C mode}. We obtain a moderately good fit with $66\%$ correlation with the original field.
The $n=3$, $m=1$ mode is picked up which represents a potential field. This may be suggestive of the
fact that the only constant $\alpha$ solution allowed within the LL modes is the potential $\alpha=0$
mode. We find that the energies of the best-fit field match that of the original field within a factor of two.
\item \textit{CK mode}. We get a fit of $64\%$ correlation with the original field. Again in this case 
the $n=3$, $m=1$ mode is picked up as in the previous case. The non-axisymmetry of this field makes it more 
 difficult to fit with a axisymmetric NLFF, which accounts for the low correlation.
\end{enumerate}

Thus, we find that we are able to get the correct configuration for the input field (as in the dipole case).
. The accuracy of the fits can be improved by taking more
grid points in our parameter search space. This is computationally extensive and will be taken up in the near future.
 We do not get good matches to linear/non axisymmetric linear force-free fields
using nonlinear axisymmetric fields where the relative helicity in these cases could not be obtained accurately:
the energy however has been obtained in all cases within a factor of two.
The morphological matches in all of the three cases are shown in Figure \ref{search}. We see that 
the overall features of fields are well captured in the fits. In this context, we remark that our fits 
for observed data are higher.

\begin{figure}[h!]
\centerline{\includegraphics[scale=0.3]{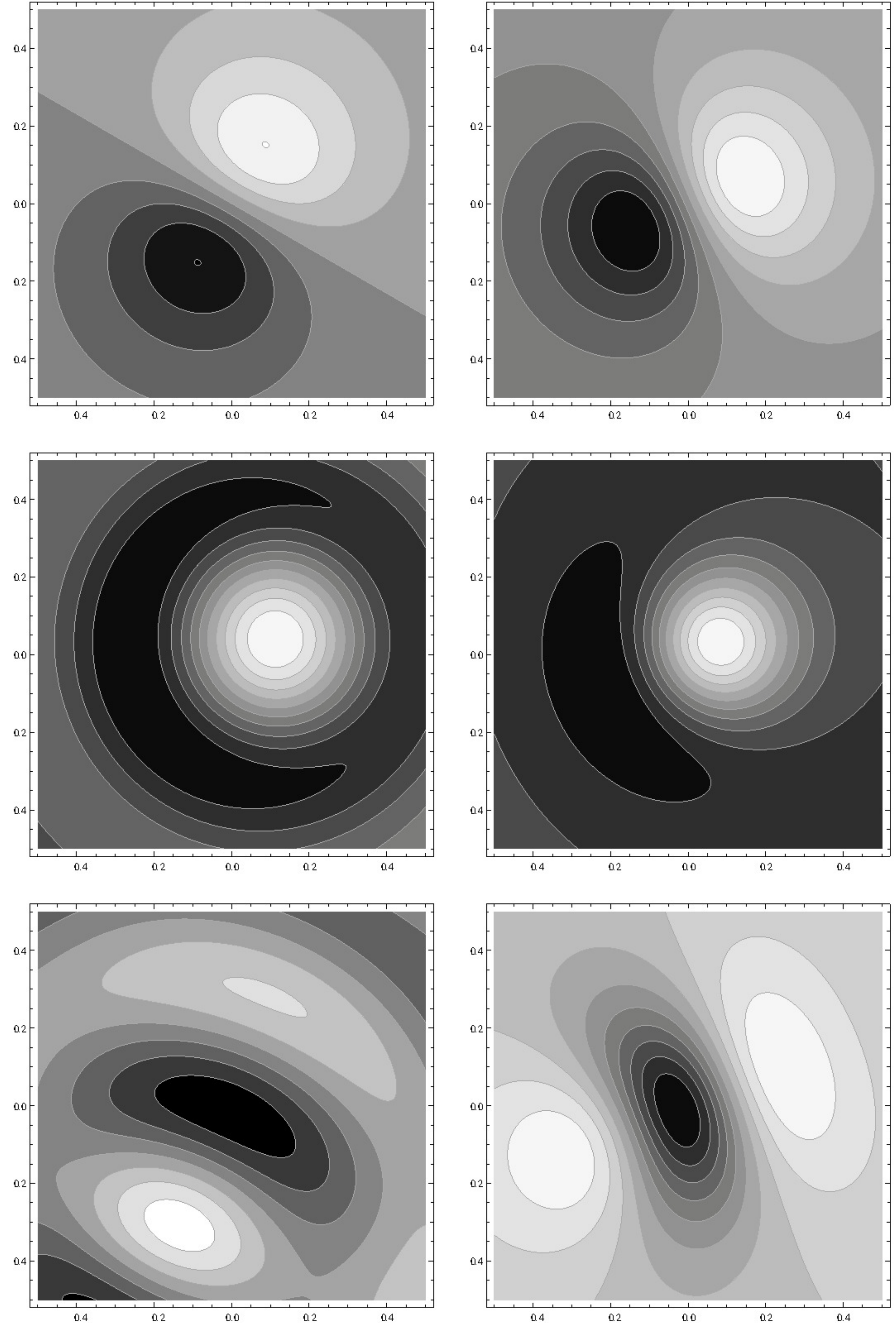}}
\caption{Left panels represent the input fields and the 
right panels represent the corresponding best-fits by LL modes. The input in the
top, middle, and bottom rows are dipole, C mode and CK mode, respectively. More details
are given in Table \ref{testtab}}.
\label{search}
\end{figure}

\section{\uppercase{Preparation of observational data}}
\label{s:prepare}
In order to compare the analytic solutions computed in this paper with the real magnetic field measurements and their associated
quantities, we use the active region magnetic field data from the spectropolarimeter (SP)
 onboard \textit{Hinode}. The (SP) is one of the instruments of the solar optical telescope (SOT). The SOT/SP obtains 
Stokes profiles with a spatial resolution of 0.3$^{\prime\prime}$ \citep{ichi} in magnetically
sensitive Fe I lines at 630.15 and 630.25 nm. 
The SOT/SP can make the map of an active region in four modes, which
are normal map, fast map, dynamics and deep magnetogram modes. 
In this study, we use the data from the fast mode, the
spatial resolution along the slit direction is 0.295$^{\prime\prime}$, 
and in the scanning direction it is 0.317$^{\prime\prime}$ pixel$^{-1}$.
The obtained Stokes profiles were calibrated using the solar software
 suites for the SP. The Stokes vectors have been inverted
using the Milne\textendash Eddington inversion \citep{skum,lites90,lites93}, and the three
components of magnetic field were obtained. The 180$^{\circ}$ ambiguity in
 the transverse field has been resolved using the
minimum-energy algorithm developed by \citet{metcalf94} and implemented 
by \citet{leka09} in Fortran. This
algorithm minimizes the electric current density and divergence simultaneously,
 selecting the field orientation with 
minimum free energy. The algorithm is the best among the several 
codes for automatically resolving the 180$^{\circ}$ ambiguity
\citep{metcalf06}. The resulting vector components have been
 transformed to the disk center \citep{ven}.
The resulting vertical, $B_z$, and transverse, $B_t$, 
field strengths have 1$\sigma$ error bars of 8 and 30 G, respectively.

We have chosen magnetograms of three active regions
 (Table \ref{tab}) spanning the years 2006 to 2007
 for our analysis.  Most of the AR appeared in the southern hemisphere at a latitude 
close to the equator.
 
\begin{table}[h!]
\label{data}
\begin{minipage}{15 cm}
 \resizebox{16cm}{2.1cm}{
\begin{tabular}{|c|c|c|c|c|c|}
 \hline
No.&  Active Region& Date and Time of Obs. & Latitude & Pixel resolution& Length, $L~(10^9)$ cm \\
 \hline
1. & &2006 Dec 12, 2000 UT && 0.306$^{\prime\prime}$&8.0   \\
 
2.&  & 2006 Dec 13, 0400 UT& & 0.306$^{\prime\prime}$&8.0  \\

3.&NOAA 10930 &2006 Dec 14, 1700 UT  &S05& 0.306$^{\prime\prime}$&8.44  \\
 
4.& &2006 Dec 14, 2200 UT  && 0.306$^{\prime\prime}$&8.44    \\
 
5.&  & 2006 Dec 15, 0545 UT& & 0.306$^{\prime\prime}$&8.44   \\
\hline
6.&NOAA 10923& 2006 Nov 11, 1430 UT&S04& 0.306$^{\prime\prime}$&8.44  \\
\hline
7.&NOAA 10933& 2007 Jan 07, 0000 UT& S05&  0.306$^{\prime\prime}$&8.44 \\
\hline
\end{tabular}
}
\caption{ Serial numbers are assigned 
to the active regions in the first column for reference. 
The date, time and latitude for the observations are given in the next two
columns. The last two columns represent the mean pixel resolution and the physical 
length scale of the magnetogram.}
\end{minipage}
\label{tab}
\end{table}

\section{\uppercase{Comparison of models to observations}}
\label{compres}
\subsection{Results}
\begin{sidewaystable*}
\begin{minipage}{235mm}
\resizebox{23.5cm}{6.75cm}{
\begin{tabular}{|c|c|c|c|c|c|c|c|c|c|c|c|c|c|}
\hline
\multicolumn{8}{|c|}{}&\multicolumn{6}{|c|}{}\\
\multicolumn{8}{|c|}{\Large{C Mode Parameters}}&\multicolumn{6}{|c|}{\Large{LL Mode Parameters}}\\
\multicolumn{8}{|c|}{}&\multicolumn{6}{|c|}{}\\
\hline 
& & & & & &  & & & &  & & &  \\ 
AR No.& {Modes}& $\alpha^{-1}/L$ & {$r_1/L,~r_2/L$}&{$\overline{E}_{ff}$  } & {$\overline{E}_{pot}$ }
 &{$\overline{E}_{free}$ }&{ $\overline{H}_{rel}$ }&  {Modes} & {$r_1/L$}&{$\overline{E}_{ff}$ } & {$\overline{E}_{pot}$ } &{$\overline{E}_{free}$ }&{ $\overline{H}_{rel}$ } \\ 
&  $n,m$& & {$\theta^\prime/~\pi,~\psi^\prime/\pi$}& {$10^{33}$ } & {$10^{33}$ } & {$10^{33}$ } & {$10^{43}$ } 
&  $n,m$& {$\theta^\prime/~\pi,~\psi^\prime/\pi$} & {$10^{33}$ } & {$10^{33}$} & {$10^{33}$ } & {$10^{43}$}  \\
& & &  &  {erg} & { erg} & { erg} & { Mx$^2$}  & &  & {erg} & { erg} & { erg} & { Mx$^2$}  \\
& & & & & &  & & & &  & & &   \\ 
\hline
& & & &   & & & &  & & &  & & \\ 
1.&1, 10& $-$0.09&1.32, 1.45&1.50&0.905&0.595 &$-$ 1.91& 3, 2&0.53&3.46&2.77&0.69&$-$0.322\\
 &  & &0.034, 1.142&  &  &   & &  &0.117, 1.37&  &  & & \\
& & & &  &  &  &  &  &  &  &  & & \\
2.&2, 10& $-$0.091&1.34, 1.85&6.27&0.760&5.51&$-$ 10.69&3, 2&0.57&11.69&9.34&2.35&$-$1.17\\
& & &$-$0.04, 0.095&  &  &  &  &  &0.67, 0.45&  &  & & \\
&  & & &  &  &  &   &  &  &  &  & & \\
3.&1, 5& $-$0.094&0.81, 0.99&2.71&1.18&1.53&$-$4.96&3, 2&0.57&5.21&4.16&1.05&$-$0.552\\
 & & & $-$0.02, 0.38 &  &  &  & &   &0.167, 1.37&  &  & & \\
& & & &  &  &  &  &   &  &  &  & & \\
4.&1, 8& $-$0.089&1.10, 1.24&1.633&0.928&0.705& $-$2.30&3,2&0.67& 4.41&3.52&0.89&$-$5.49\\
& & & 0.015, 1.25&    &  & & &  &0.117, 1.29 &  &  & & \\
&  &   &  &  &  & & &  &  &  &  & & \\
5. & 1, 8& $-$0.089&1.10, 1.24&1.54&0.877&0.663& $-$2.18&3, 2&0.67&4.16&3.32&0.84&$-$0.518\\
& & &0.015, 1.25&    &  & & &  &0.117, 1.29 &  &  & & \\
&  & &   &  &  & & &  &  &  &  & & \\
\hline
& & & &   & & & &  & & &  & & \\ 
6.&1, 0&$-$0.461&1.65, 2.07&119.4&90.5&28.9&$-$535.3&1, 1 &0.43&10.70&10.70&0.0&0.0\\
&  & &$-$0.97, 0.75&  &    & & &  &0.83, 1.33&  &  & & \\
& & & &    &  & & &  &  &  &  & & \\
\hline
& & &  &  & & & &  & & &  & & \\ 
7.&1, 0&$-$0.75&2.95, 3.37&178.1&161.1&17&$-$530.2&3, 1&0.57&2.063&2.063&0.0&0.0\\
& & & $-$1.0,  $-$0.15&  &    & & &  &$3.18\times 10^{-7}$, 1.67&  &  & & \\
& & &   &  &  & & &  &  &  &  & & \\
\hline
\end{tabular}
}
\caption{The table contains the correlations of the simulated and
observed magnetograms for their best-fit with C and LL modes with
the corresponding parameters and the estimations of the free energy
and relative helicity. Details of the different active regions
for each row are given in the corresponding rows in Table \ref{tab}. The numbers $(n,m)$ refer to the
radial and angular modes, and $\alpha$ is the proportionality
constant between $\mathbf{B}$ and $\mathbf{J}$, which is scaled with respect
to $L$, the length scale  of the magnetogram. The inner radius is represented by $r_1$ 
where the magnetogram is placed and $r_2$ is the outer radius of the computation region.
 In the case of C modes, $r_2$ is finite because we have restricted the calculations to only one radial 
oscillation, whereas for LL modes, $r_2$ is infinite because the fields tend
to zero only at infinity. The Euler angles through which the magnetogram is rotated are 
represented by $\theta^\prime$ and $\psi^\prime$.
$\overline{E}_{ff}$, $\overline{E}_{pot}$, $\overline{E}_{free}$ and $\overline{H}_{rel}$
 are the normalized force-free energy, potential energy, free energy and the relative helicity 
of the magnetic field configuration, respectively.}
\end{minipage}
\label{results}
\end{sidewaystable*}
We list our findings below:
\begin{enumerate}
\item All of the field configurations analyzed were found 
to be negatively twisted as seen from the $\alpha$ for
the C modes and the sign of the helicity for the LL modes.
 The fits with nonlinear LL modes are substantially better
than the linear C modes, confirming the nonlinear nature of the force-free fields.

\item Table \ref{corrtable} lists all of the parameters compiled for the AR listed in Table \ref{tab}.

\item The AR 10930 has been fit by LL modes with a figure of merit $c=0.7$-$0.8$ and $d=0.65$-$0.7$.  The energy ratio,
$\epsilon$, is away from unity due to the fact that the data is noisy
 and that we had not done any preprocessing. 
There was an X3.4-class flare on 2006  December 13, and we confirm
 in both models a substantial decrease in free energy and relative helicity after the flare.
The relative helicity and free energy in the C mode increased and in the LL mode decreased
marginally after the X1.5-class flare on 2006 December 14.

\item The two ARs 10923 and 10933 with single-polarity fitted with potential fields with a high figure of merit $>90\%$.
They also show good correlation numbers for $d$ and $\epsilon$ (near unity).

\item 
The formula (\ref{np}) for the predicted $m_p(n_p, r_1/L)$
 bears out for the force-free configurations
 found for AR 10930 (see first five rows of
 Table \ref{results} for the modes $(n,m)$ for C and LL modes).
 For example, in the case of C modes, 
$m_p= 2m-2=\{18, 18, 8, 14, 14 \}$ for the five cases and we 
find the estimated $m_p$ from Equation (\ref{np}) to be $\{16, 16, 7.25, 11.7, 11.7 \}$. 
For the five cases of LL, $m_p= n+m-1=$ 4 in all of the five cases, whereas the estimated 
$m_p=\{4.25,4.6,4.6, 3.8, 3.8\}$ from Equation (\ref{np}).
 It is clear that because the $m_p$ estimates are
closer for the LL modes than for the C modes, the corresponding figures
of merit are higher for the C modes. The suggested $m_p$ 
estimates are for the lowest energy configurations.

\end{enumerate}
\begin{figure}[h!]
\centerline{\includegraphics[scale=0.25]{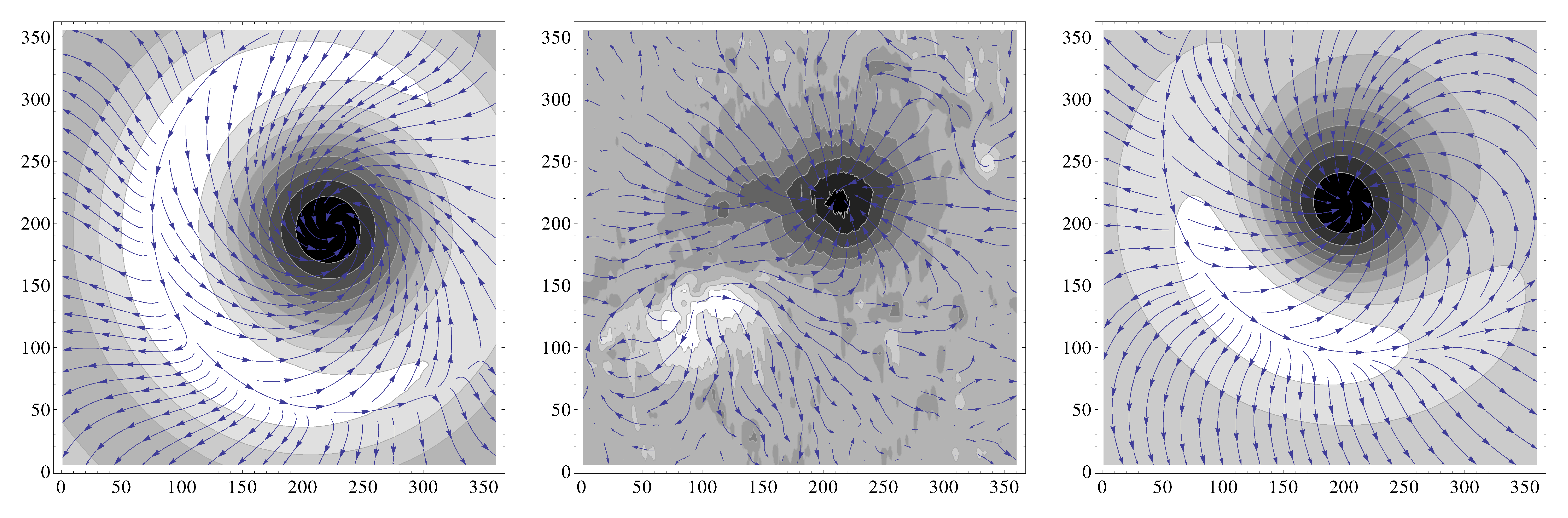}}
\centerline{\includegraphics[scale=0.25]{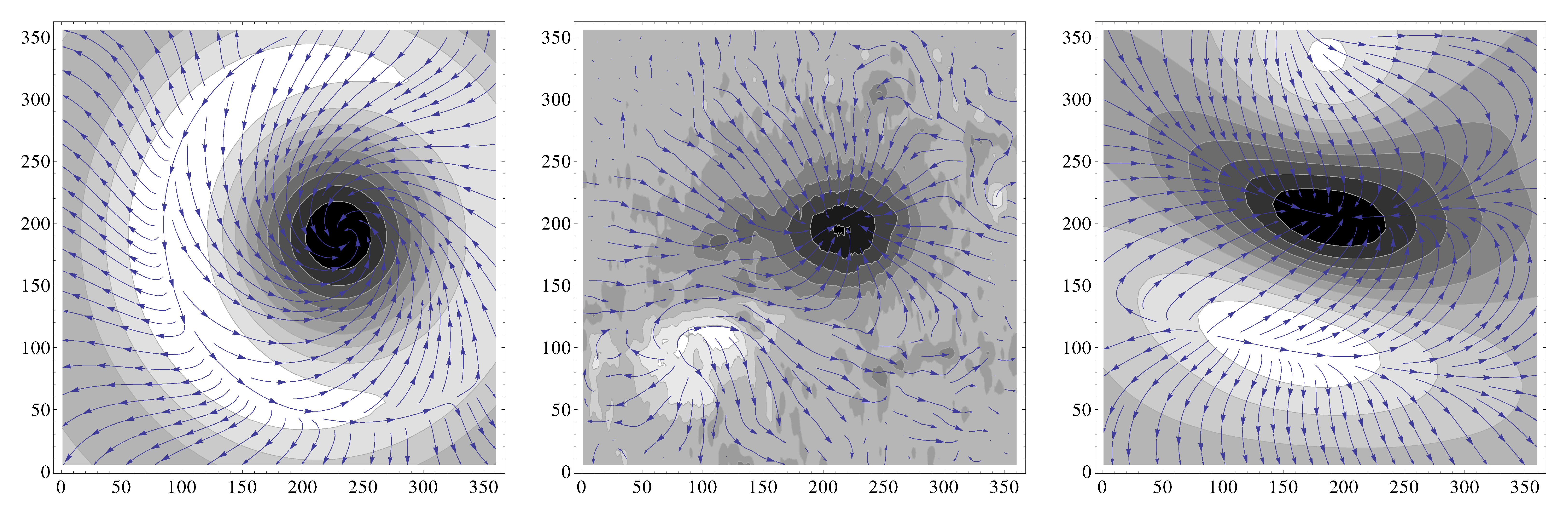}}
\caption{Magnetograms for active region NOAA 10930 are shown in the top and bottom panels of the 
figure for the dates 2006 December 12 and 13, respectively.
 The left and right panels represent the 
magnetograms simulated by the C and LL modes, respectively.
 The middle panel represents the magnetogram
observation by \textit{Hinode}.
 There was an occurrence of an X 3.4-class flare between the dates; the
figure depicts the field configuration before and after the flare.}
\end{figure}
\begin{figure}[h!]
\centerline{\includegraphics[scale=0.25]{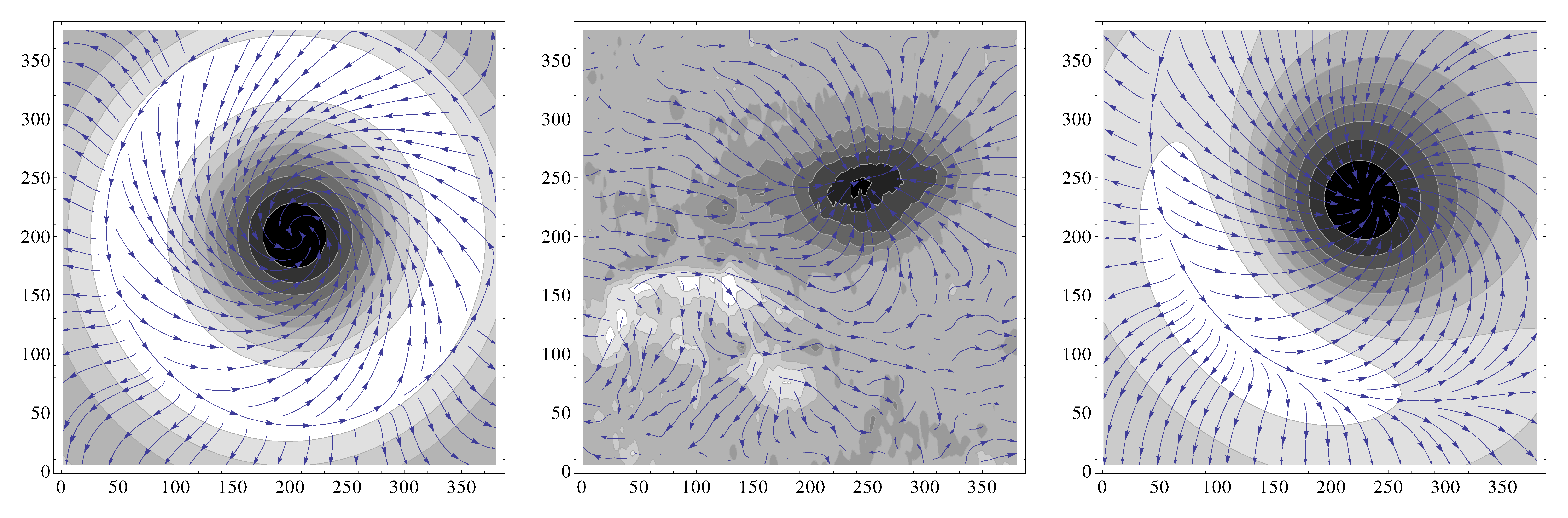}}
\centerline{\includegraphics[scale=0.25]{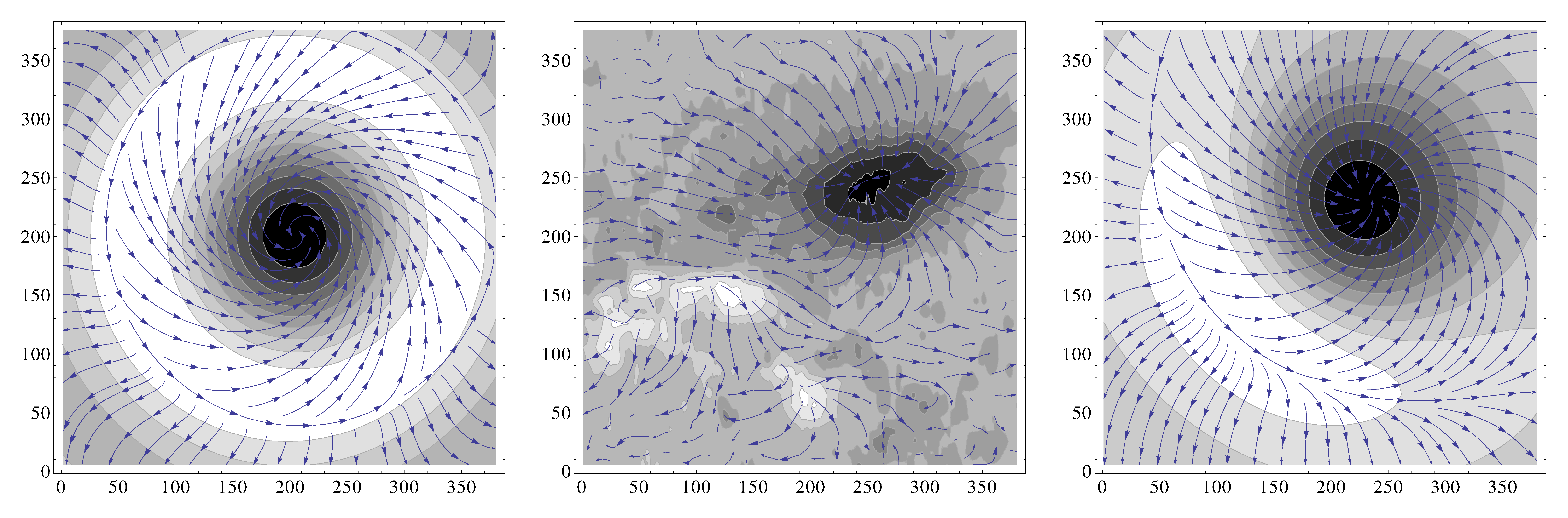}}
\caption{Magnetograms for active region NOAA 10930 are shown in the top and bottom panels of the 
figure for the dates 2006 December 14 and 15, respectively.
 The left and right panels represent the 
magnetograms simulated by the C and LL modes, respectively. 
The middle panel represents the magnetogram
observation by \textit{Hinode}. 
There was an occurrence of an X-1.5 class flare between the dates; the
figure depicts the field configuration before and after the flare.}
\end{figure}
\begin{figure}[h!]
\centerline{\includegraphics[scale=0.6]{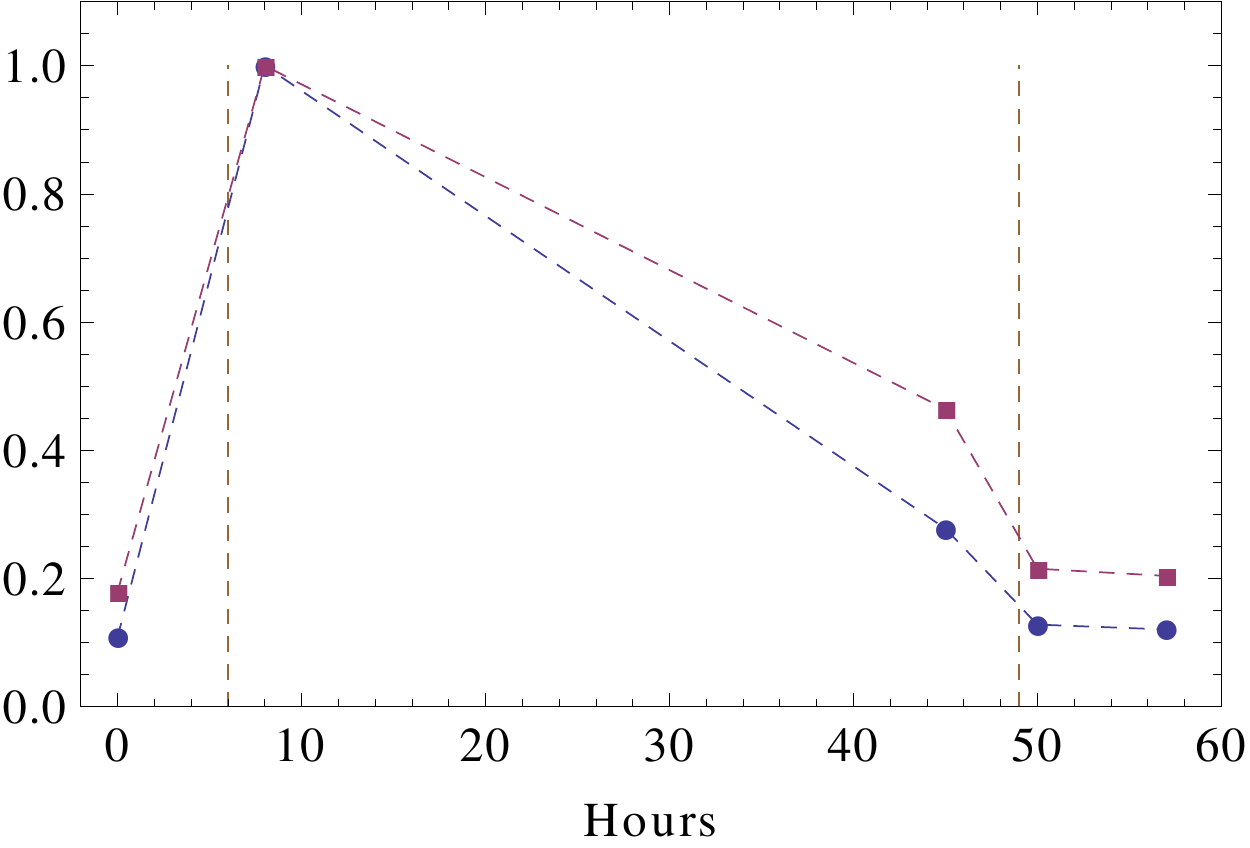}\hskip .5 cm \includegraphics[scale=0.6]{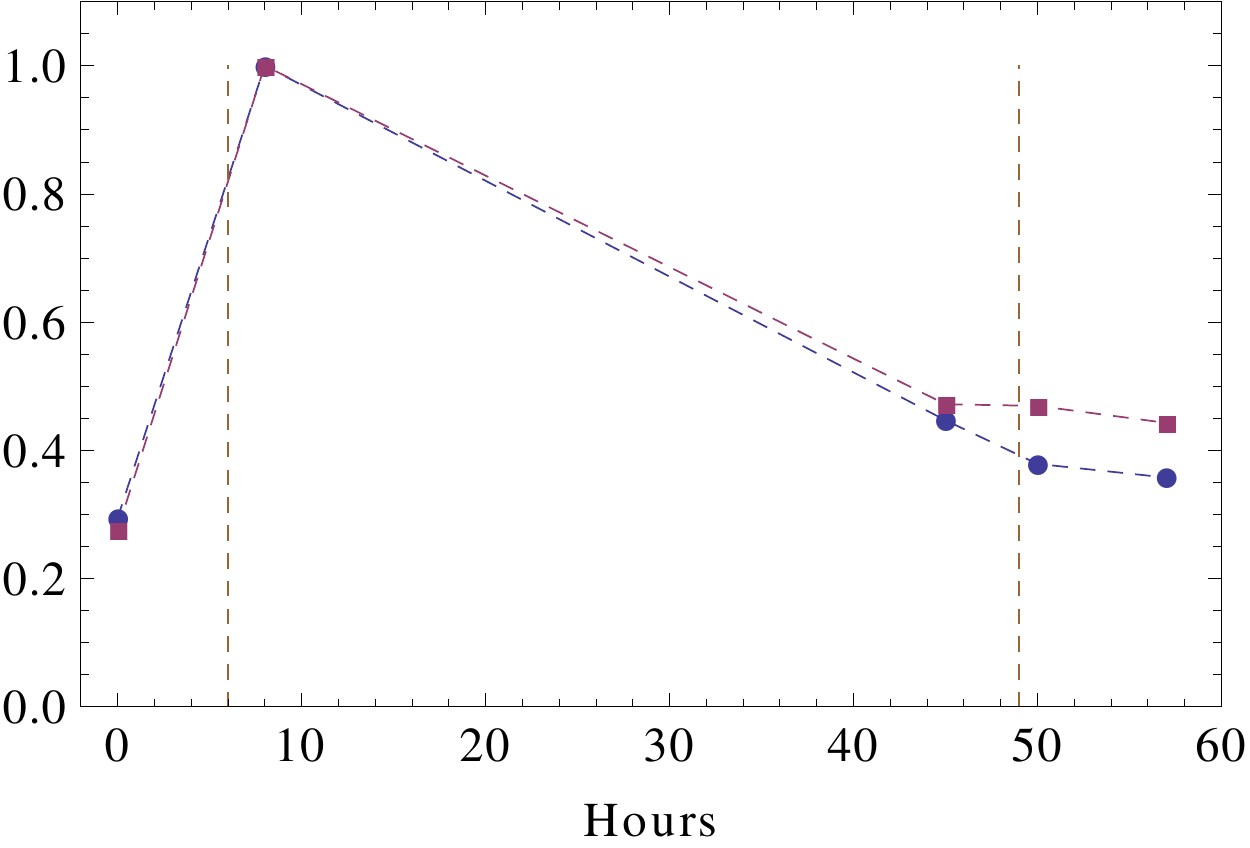}}
\caption{Free energy (circle) and relative magnetic helicity (square) simulated for AR 10930 
(see rows 1-5 in Table \ref{results}) for the C (left) and LL (right) modes, 
plotted as a function of time. The time of the first observation (2006 December 12;
 UT 2000) is set as 0 hr. Them vertical lines represent the times of X3.4- and X1.3-class flares,
respectively. The values on the y-axis have been normalized with respect
to their maximum.}
\label{f:time}
\end{figure}

\begin{table}
\begin{tabular}{|c|c|c|c|c|c|c|c|}
\hline
\multicolumn{5}{|c|}{}&\multicolumn{3}{|c|}{}\\
\multicolumn{5}{|c|}{\Large{C Mode}}&\multicolumn{3}{|c|}{\Large{LL Mode}}\\
\multicolumn{5}{|c|}{}&\multicolumn{3}{|c|}{}\\
\hline 
& & & & &  & &  \\ 
N0.& AR No.&
$c$& $d$&$\epsilon$&$c$ & $d$ &$\epsilon$\\
& & & &  &  & &  \\ 
\hline
1.& &0.58&0.521& 1.834&0.81&0.695&1.273\\
 & & & & &  &  &  \\

2.& &0.581&0.511& 1.876&0.70&0.645&1.88\\
 & & & & &  &  &   \\

3.&NOAA 10930 &0.33&0.324& 1.791&0.75&0.673&1.687\\
 & & & & &  &  &   \\

4.& &0.42&0.395& 1.724&0.76&0.698&1.765\\
 & & & & &  &  &   \\

5.& &0.40&0.374& 1.644&0.74&0.697&1.682\\
 & & & & &  &  &   \\
\hline
6.&NOAA 10923&0.76&0.888& 1.517&0.92&0.928&0.943\\
 & & & & &  &  &   \\
\hline
7.&NOAA 10933&0.56&0.788& 2.40&0.95&0.871&1.113\\
 & & & &  & & &   \\
\hline
\end{tabular}
\caption{The correlation parameters obtained for estimating the
goodness of fit for different active regions. The definitions of
 the correlation parameters are given in Section \ref{fit}.}
\label{corrtable}
\end{table}
\voffset 0 in
\subsection{Discussion of the Results}

AR 10923 and AR 10933 show good fits ($c > 90\%$) with single-polarity potential configurations and 
are negatively twisted with the energies of $10.7 \times 10^{33}$
and $2.063 \times 10^{33}$ erg, respectively. The corresponding goodness of fit
parameters ($d, \epsilon$) are near unity and indicate good fits (see Table \ref{corrtable}).
 
Active region NOAA 10930 is a center of focus for several studies and the \textit{Hinode}/SOT has
followed the active region for several days in many wavelength regions.
 The SP produced vector magnetograms of this region until it disappeared on the west limb of the Sun. Using the vector
magnetic field measurements at the photospheric levels and by applying a technique called preprocessing
several authors employ NLFF extrapolation methods to compute the coronal magnetic fields 
(e.g., Schrijver et al. 2008). An X3.4-class flare has
occurred in this active region on 2006 December 13. Using the 3D
magnetic fields information, Schrijver et al. (2008) found 
3$\times$10$^{32}$~ergs of drop in free energy after the 
flare compared to the preflare free energy. On the other hand, Guo et~al. (2008)
find that only 2.4$\times$10$^{31}$~erg of energy were released during the flare. At the same time,
using a similar technique, Jing et~al. (2010) did not
 find any release in free energy during the flare; instead
they found a slight increase in free energy after the flare.
\citet{he2011} also estimate the free energy for three time series
 vector magnetograms of the same solar active region,
 NOAA 10930 through NLFFF extrapolation which
 were observed in a 26 hour period from 2006 December 10-11  .
 They note a rise in the free energy 
of the system from 1.25-1.42 $10^{33}$ erg.
 These results are summarized in Table \ref{compare}
 along with our results. 
\begin{table}[h!]
\resizebox{18cm}{2.4cm}{
\begin{tabular}{|c|c|c|c|c|c|c|c|c|c|c|c|c|}
\hline
\multicolumn{1}{|c|}{}&\multicolumn{5}{|c|}{}&\multicolumn{5}{|c|}{}&\multicolumn{1}{|c|}{}&\multicolumn{1}{|c|}{}\\
\multicolumn{1}{|c|}{\Large{Model$^{ref}$}}&\multicolumn{5}{|c|}{\Large{Pre-flare}}
&\multicolumn{5}{|c|}{\Large{Post-flare}}&\multicolumn{1}{|c|}{\Large{$\Delta \overline{E}_{free}$}}
&\multicolumn{1}{|c|}{\Large{$\Delta \overline{H}_{rel}$}}\\
\multicolumn{1}{|c|}{}&\multicolumn{5}{|c|}{}&\multicolumn{5}{|c|}{}&\multicolumn{1}{|c|}{}&\multicolumn{1}{|c|}{}\\
\hline 
 & & & & & &  & & & & & & \\ 
& $\overline{E}_{ff}$ & $\overline{E}_{pot}$&$\overline{E}_{ff}/\overline{E}_{pot}$ & $\overline{E}_{free}$&$\overline{H}_{rel}$ &$\overline{E}_{ff}$ & $\overline{E}_{pot}$ &$\overline{E}_{ff}/\overline{E}_{pot}$&$\overline{E}_{free}$&$\overline{H}_{rel}$& & \\ 
  &$10^{33}$  &$10^{33}$ & & $10^{33}$& $10^{43}$  & $10^{33}$  & $10^{33}$  & &$10^{33}$& $10^{43}$ &$10^{33}$& $10^{43}$ \\
 &erg & erg & & erg &Mx$^{2}$&  erg &  erg & &erg &Mx$^{2}$&  erg&Mx$^{2}$ \\
\hline
C modes$^a$ &1.5 &0.905 &1.66 &0.595 & $-$ 1.91 &6.27 &0.76 &8.25 &5.51 &$-$10.69& 4.915&$-$8.78  \\ 
\hline
LL modes$^a$ &3.46 &2.77 &1.249 & 0.69&$-$0.322& 11.69& 9.34&1.252& 2.35&$-$1.17& 1.66&$-$0.848  \\
\hline
Current-field iteration $^b$ &\nodata &\nodata & 1.32&\nodata & \nodata&\nodata &\nodata &1.14 &\nodata & \nodata&$-$ .32 &\nodata \\ 
\hline
Optimization$^c$&1.33 &1.2 &1.13 &1.11 &\nodata &1.27 &1.16 &.11 &1.09 &\nodata &$-$ .02&\nodata \\ 
\hline
Weighted optimization$^d$ &\nodata & \nodata&\nodata &$\sim$.75 & \nodata&\nodata &\nodata &\nodata &$\sim$.85&\nodata &.1 &\nodata \\
\hline
\end{tabular}
}
\caption{The values for the energy of the force-free field and the corresponding potential field for the active region
NOAA 10930 mentioned in literature are compiled in the table along with our results for reference. The quantities
such as the free energy of the configuration, ratio between energies of the force-free and potential field and the change 
 in free energy before and after the flare are also mentioned. a: current paper, b: Schrijver et. al. (2008)
, c: Guo et. al. (2008), d: Jing et. al. (2010)}
\label{compare}
\end{table}

We were able to get good correlations for this region for both C (40-60 \%) and LL (70-80\%) modes;
 see Figures 7 \& 8 and Table 4.
We find that free energies derived from the LL model are consistent with a drop after 
both of the flare events (note that the time coverage before and after the flare is not complete),
indicating a strong probability of a peak in the free energy (and relative helicity) 
just before the first flare event, see Figure \ref{f:time}.
This picture is conducive to the idea that the loss of the free energy in the photosphere is strongly related
to the energy dissipated in the flare events. 

The coronal mass ejection (CME) associated with this event carried kinetic energy
(deprojected velocity ) of 4.5$\times$10$^{32}$~erg with it \citep{ravindra10}.
 This is in rough agreement with our estimate of a loss of 1.66$\times$10$^{33}$~erg
 (LL modes) because about half of this
would be released in the kinetic energy channel. The magnetic cloud associated with the CME had a
helicity of about -7$\times$10$^{41}$~Mx$^{2}$ as estimated here by
 \citet{ravindra11} which is much
less than the drop in helicity estimated for the LL modes to be -0.322
 $\times$10$^{43}$~Mx$^{2}$.
However our estimate of the relative helicity injected into the active region corona 
is found to be -0.848$\times$10$^{43}$~Mx$^{2}$ before the initiation of an X3.4-class
flare which is comparable to the $-4.3\times$10$^{43}$~Mx$^{2}$ found by Park et al. (2010). 
All of the NLFF extrapolation techniques employed (other than in this work) in computing the free energy
using the vector magnetic field data made use of preprocessing technique
 to make the field closer to force-free. However, in doing so the field gets smoothed 
thereby reducing the field strength and hence the free energy.  

For the same data sets other authors using different NLFF extrapolation techniques for the 
analysis obtain slightly different results. In some cases, there is an increase in the
free energy after the flare, whereas in other cases they find that it decreases.
In our analysis, both the free energy and relative helicity increase after the first 
flare, as can be seen from the last two columns in Table \ref{compare}. The ratio of the energies 
of the 	force-free field and the potential field remains almost constant before and after
 the flare for the LL modem, whereas for the C mode it increases.
 This can be used to infer that during the flare process there is a dynamic evolution of both 
the force-free field and the potential field from a lower to higher energy state,
implying that there was a peak in free energy and relative helicity between the two observations which are separated 
by a large time gap of 8 hr.

\section{\uppercase{Summary and conclusions}}
\label{s:summary}
 \begin{table}
\resizebox{18cm}{9cm}{
\begin{tabular}{|l|}
\hline
\centerline{\bf C MODES}\\ \hline
\\
$
\mathbf{B}(r_1<r<r_2)=\left(\frac{-J_{m+3/2}(\alpha r)}{r^{3/2}}\frac{d }
{d \mu}[(1-\mu^2)C_m^{3/2}(\mu)],\frac{-1}{r}\frac{d }
{d r}[r^{1/2}J_{m+3/2}(\alpha r)](1-\mu^2)^{1/2}C_m^{3/2}(\mu),
\frac{\alpha J_{m+3/2}(\alpha r)}{r^{1/2}}(1-\mu^2)^{1/2}C_m^{3/2}(\mu)\right)
$
\\
\\
$
\mathbf{A}(r_1<r<r_2)=\mathbf{B}/\alpha;\quad
a_{m+1}=\frac{(m+2)r_1^{m+3/2} J_{m+3/2}(\alpha r_1)}{ r_1^{2m+3}-r_2^{2m+3}};\quad
b_{m+1}=\frac{(m+1)r_2^{2m+3} r_1^{m+3/2} J_{m+3/2}(\alpha r_1)}{ r_1^{2m+3}-r_2^{2m+3}}
$
\\
\\ 
$
\mathbf{B}_P(r_1<r<r_2)=\left(\left[(m+1) a_{m+1} r^{m}-\frac{(m+2)b_{m+1}}{r^{m+3}}\right]P_{m+1}(\mu),\right.
\left.-(1-\mu^2)^{1/2}\left[ a_{m+1} r^{m}+\frac{b_{m+1}}{r^{m+3}}\right]\frac{dP_{m+1}}{d\mu},0\right)
$
\\
\\ 
$
\mathbf{A}_P(r_1<r<r_2)=\left(0,0,(1-\mu^2)^{1/2} P^\prime_l(\mu)\left[\frac{ a_l r^l}{l+1}-\frac{b_l}{lr^{l+1}}\right]\right)
$
\\
\\ 
$
E_v(r)=\frac{(m+1)(m+2)}{2(2m+3)}\left [r\left [\frac{d}{d r}
\left\{r^{1/2} J_{m+3/2}(\alpha r) \right\}\right]^2+\left\{\alpha^2r^2-(m+1)(m+2)
\right\}J^2_{m+3/2}(\alpha r)\right]; 
$
\\
\\
$
E_{\mathrm{ff}}(\alpha,n,m, r_1, r_2)=E_v(r_2)-E_v(r_1)
=\frac{(m+1)(m+2)}{2(2m+3)}\Bigl 
 [2\int_{r_1}^{r_2}\alpha^2r J^2_{m+3/2}(\alpha r)dr
-r_1^{1/2} J_{m+3/2}(\alpha r_1)\frac{d}
{d r}\{r^{1/2}J_{m+3/2}(\alpha r)\}|_{r=r_1}\Bigr]
$
\\
\\

$
E_{pot}(m,r_1,r_2)=\frac{1}{2(2m+3)}\int_{r_1}^{r_2}\Bigl[\left((m+1)a_{m+1}r^{m+1}-
\frac{(m+2)b_{m+1}}{r^{m+2}}\right)^2
+(m+1)(m+2)\left(a_{m+1}r^{m+1}+\frac{b_{m+1}}{r^{m+2}}\right)^2\Bigr]dr
$
\\
\\ 
$
H_{rel}^{FA}(\alpha,n,m,r_1,r_2)
 =\frac{8\pi E_{\mathrm{ff}}}{\alpha}+\frac{4\pi(m+1)(m+2)}{\alpha (2m+3)}\Bigl[\alpha^2 
\int_{r_1}^{r_2}\left(\frac{a_{m+1}r^{m+1}}{m+2} -\frac{b_{m+1}}{(m+1)r^{m+2}}\right )r^{3/2}J_{m+3/2}
(\alpha r) d r 
$
\\
$~~~~~~~~~~~~~~~~~~~~~~~~~~~
+r_1^{1/2}\left(a_{m+1}r_1^{m+1}+\frac{b_{m+1}}{r_1^{m+2}}\right)J_{m+3/2}(\alpha r_1) \Bigr]
$
 \\
\\
$
H_{rel}^{B}(\alpha,n,m,r_1,r_2)=\frac{8\pi\alpha(m+1)(m+2)}{2m+3}\int_{r_1}^{r_2}rJ_{m+3/2}^2(\alpha r)dr.
$
\\
\\
\hline
\centerline{\bf LL MODES} \\ \hline
$
\mathbf{B}(r<r_2)= \left(\frac{-1}{r^{n+2}}\frac{dP}{\partial\mu},
\frac{n}{r^{n+2}}\frac{P}{(1-\mu^2)^{1/2}},\frac{a}{r^{n+2}}\frac{P^{(n+1)/n}}{(1-\mu^2)^{1/2}}\right);
\quad
\mathbf{A}(r<r_2)=\left(0,\frac{-a}{n r^{n+1}}\frac{P(\mu)^{(n+1)/n}}{(1-\mu^2)^{1/2}},\frac{1}{r^{n+1}}\frac{P(\mu)}{(1-\mu^2)^{1/2}}\right)
$
\\
\\
$
a_l=0,\quad b_l=\frac{2l+1}{2(l+1)}r_1^{l-n}\int_{-1}^1\frac{dP}{d\mu}P_l(\mu)d\mu;
\quad
\mathbf{B}_P(r_1<r<r_2)=\left(\sum_{l=0}^{\infty}-(l+1)\frac{b_l}{r^{l+2}}P_l(\mu),\sum_{l=0}^{\infty}\frac{-b_l}{r^{l+2}}
(1-\mu^2)^{1/2}\frac{dP_l}{d\mu},0\right).
$
 \\
\\ 
$
\mathbf{A}_P(r_1<r<r_2)=\left(0,0,(1-\mu^2)^{1/2} P^\prime_l(\mu)\left[\frac{ a_l r^l}{l+1}-\frac{b_l}{lr^{l+1}}\right]\right);
\quad
E_{pot}(l,r_1)=\sum_{l=0}^\infty\frac{b_l^2(l+1)}{2(2l+1)r_1^{2l+1}}
$
\\
\\
$
E_{\mathrm{ff}}(n,m, r_1)=\frac{1}{4(2n+1)r_1^{2n+1}}\int_{-1}^1d\mu
\left[ P^\prime(\mu)^2+\frac{n^2 P(\mu)^2}{1-\mu^2}+\frac{a^2 P(\mu)^{(2n+2)/n}}{1-\mu^2}\right]
=\frac{1}{4 r_1^{2n+1}}\int_{-1}^1\left\{\left(\frac{d P}{d \mu} \right )^2
-\frac{(n^2+a^2 P^{2/n})P^2}{(1-\mu^2)}\right\}d \mu
$
\\
\\
$
H_{rel}^{FA}(n,m,r_1)=-2\pi a \sum_{l=0}^\infty \int_{-1}^1  \frac{b_l}{n l r_1^{n+l}}P^{1+1/n} \frac{d P_l}
{d \mu}d\mu
=H_{rel}^{B}(n,m,r_1)=\frac{2\pi a}{n r_1^{2n}} \int_{-1}^1  \frac {P^{2+1/n}}{(1-\mu^2)}d\mu. 
$
\\
\\
\hline
\end{tabular}
}
\caption{Formulary for the various quantities calculated for the C and LL modes. 
$\mathbf{B}$ and $\mathbf{A}$ denote the force-free magnetic field and its corresponding 
vector potential. The same quantities for the potential field are denoted by $\mathbf{B}_P$
and $\mathbf{A}_P$ respectively. $E_{\mathrm{ff}}$, $E_{pot}$, $E_{\mathrm{free}}$ and $H_{rel}$
are the force-free energy, potential energy, free energy and the relative helicity 
of the magnetic field configuration respectively calculated using the Finn Antonesen 
\& Berger formulae that are analytically equivalent.}
\label{t:formulae}
\end{table}
Here we first summarize the key results of this paper.
\begin{enumerate}
  \item 
{\em Analytic Results.}

We have shown that there are two solutions possible (albeit known already and denoted here as C and LL)
from the separability assumption. We calculate the energies and relative helicity of the allowed force-free
fields in a shell geometry.
The final expression for the field of C modes is given in Equation (\ref{bchand}). 
We then calculated the corresponding potential field for calculating relative helicity in this region. 
The expressions for the potential field and its vector potential are given 
in Equations (\ref{bpchand1}) 
and (\ref{Apchand1}). The relative helicity thus calculated is given by Equation (\ref{hrelchand}).
The expression for energies of the force-free field and the potential field 
are given by Equations (\ref{eint}) and (\ref{epotchand}), respectively whereby we can calculate the free energy of the 
system using Equation (\ref{efree}).
The alternative expressions for the energy of the force-free field and relative
helicity are given in Equations (\ref{evcmode}) and (\ref{hrelcb}), respectively which are analytically in agreement with the 
previous expressions.
 
For the LL mode we were able to extend the solution set obtained in \citet{low90} from $n=1$ to  
all rational values of $n=\displaystyle{\frac{p}{q}}$ by solving the Equation (\ref{feq}) for all 
cases of odd $p$ and for cases of $q>p$ for even $p$, in effect extending solution to practically all $n$. 
The final expression for the magnetic field is given by Equation (\ref{flow}) 
and its vector potential by 
Equation (\ref{alow1}). The expression for the potential field consistent with 
this force-free field is given by Equation (\ref{Bplow1}) and the corresponding vector potential is given by 
Equation (\ref{Apchand1}) with the constants evaluated from Equation (\ref{alLL1}). The relative helicity in the 
region using the Finn\textendash Antonsen formula is given by Equation (\ref{hrellow}). The energies for the force-free 
and potential fields are given by Equations (\ref{efflow}) and (\ref{epotlow}), respectively. 
Again, the alternative expressions for the energy of the force-free field and relative
helicity are given in Equations (\ref{effvll}) and (\ref{hrbll}),
respectively which are analytically in agreement with the previous expressions.
For convenience these results are included in the formularies for 
C and LL modes in Table \ref{t:formulae}.

\item 
{\em Numerical Results.}

We formulated a search strategy with parameters including two Euler rotations of the force-free sphere 
and a variable set that corresponds to the various C and LL modes; see Section \ref{search}.
A study of effectiveness of our search strategy is presented in Section\ref{effective}. Here we find that
we are able to get the correct configuration for the input field (as in the dipole case)
and are able to fit the energies within a factor of two.
 We then studied the field configurations for three active regions, (Table \ref{tab}) and calculated the free energy and relative
helicity for these cases. We were able to get reasonable fits for the above cases; see Table \ref{corrtable}.
All of the field configurations analyzed were found to be negatively twisted as seen 
from the $\alpha$ for
the C modes and the helicity of the LL modes; see Table \ref{results}. 
The fits with nonlinear LL modes seem to 
be better than the linear C modes. In the case of AR 10930, there was an X3.4-class
 flare on 2006 December 13, 
and we confirm in both modes a substantial decrease in free energy and 
relative helicity after the flare.
A comparison of results obtained in this paper with those in the literature
 for the same flare event is presented
in Table \ref{compare}. The relative helicity and free energy in the C mode increased and 
in the LL mode decreased marginally after the X1.5-class flare on 2006 December 14. 
The two ARs 10923 and 10933 with single-polarity show very high correlation ($>90\%$) 
with potential fields.
We were not able to explore the full parameter space because 
of the computational constraints 
mentioned in Section \ref{search}. Because our best-fit with the 
observational data for the LL modes is
substantially better (~75\%) than those obtained in the test cases,
this lends much credibility to the results presented in the paper. 
\end{enumerate}
 
We find that the approach taken here is fairly good in estimating the quantities of
interest, namely relative helicity and free energy; see Table \ref{compare} and Section \ref{compres}.
In order to compare with the other estimates available in the literature,
 where the potential fields
are usually extended from the planar surface of the magnetogram to a cuboidal
volume over the magnetogram, we rescale our physical quantities obtained for a hemisphere
by the factor of the solid angle subtended by the magnetogram at the center. 
This enables us to approximate their trend before and after a flare event. 
The validity of this approximation can be seen from the general agreement with 
other estimates (including observations). An advantage
in this method is its ease and utility in calculating these physical quantities, in particular
the relative helicity is thus far not calculated by other approaches. Further,
 we don't have to assume any other boundary conditions for the side walls, as  required
in the other extrapolation techniques using the cuboidal volume.
 This method can also provide a useful reconstruction of the NLFFs 
as well as reasonable input field for other numerical techniques.
It is clear that nonlinear LL modes are dominantly better fits than the linear C modes. 
The search is now limited by computational constraints; in the future,
we hope to improve the fits by applying the method to a larger space of geometrical parameters
and in more cases of mode numbers $n$ and $m$.

The LL solutions of $n=1$ in \citet{low90} and $n=5,7,9$ (odd cases) in \citet{flyer04}
have been extended here to the cases of nearly all $n$.  
 The topological properties of these extended solutions can be further studied by considering other
 boundary conditions.The analytic solutions for LL suffer from the problem of a singularity at 
the origin, which render them unphysical; this implies that more realistic boundary
conditions are necessary.

To learn more about the evolution and genesis of these structures, it would be useful to carry
 out dynamical simulations that allow for footpoint motions with the analytic input fields constructed above 
to study how the nonlinearity develops; a stability analysis of the nonlinear modes would also be a useful tool 
(\citet{berger85} has analyzed the linear constant $\alpha$ case). Clearly, these are difficult mathematical 
problems to be addressed in the future.
 
\acknowledgements
The authors thank the referee for the valuable, constructive and insightful comments.
The authors would like to acknowledge the \textit{Hinode} team for the photospheric magnetogram data. 
\textit{Hinode} is a Japanese mission developed and launched by ISAS/JAXA, with NAOJ as domestic partner and 
NASA and STFC (UK) as international partners. It is operated by these agencies in cooperation 
with ESA and NSC (Norway). We would also like to acknowledge IIA for providing the computational facilities 
and Sandra Rajiva for editorial support. A.P. would like to thank CSIR for the SPM fellowship.

\appendix
\section{\uppercase{Boundary conditions and formulae for C mode fields}}
\label{cmode}
\begin{figure}[h!]
\centerline{\includegraphics[scale=.35]{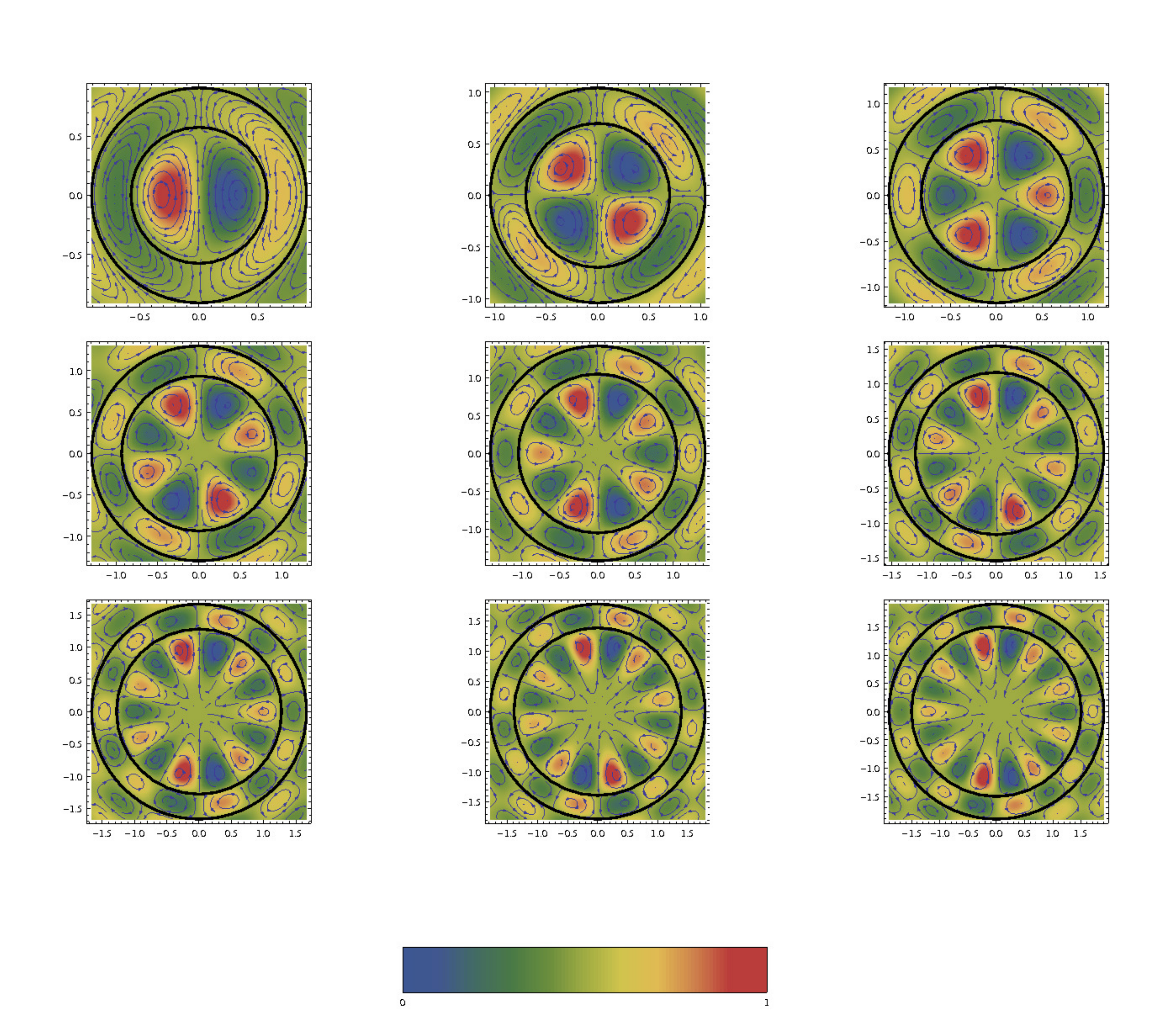}}
\caption{Different angular modes from $m$ = 1 (top left) to $m$ = 9 (bottom right)
are shown. The contours represent the poloidal stream function $\psi$
and the density plot represents the strength of the azimuthal field $B_{\phi}$.
The two circles are drawn at first and second radial roots. The 0 and 1 in the 
legend corresponds to the minimum and maximum values of $B_{\phi}$, respectively.}
\label{cplt}
\end{figure}

The conditions to be satisfied at a interface (where $\alpha$ has a discontinuity) are
\begin{enumerate}
 \item  The divergence condition implies the continuity of the normal 
component of the magnetic field, whereas the absence of surface currents 
on the boundaries leads to continuity of the parallel components of the field. Therefore magnetic field $\mathbf{B}$ should be continuous.
\item The normal component of the current density $\mathbf{J}$ should be 
continuous because there is no accumulation of charges at the boundary. 
\end{enumerate}
The second condition requires the continuity of  the normal 
component of $\alpha\mathbf{B}$; in contrast, the first condition requires the 
normal component of $\mathbf{B}$
to be continuous. If $\alpha$ changes discontinuously at a spherical shell (say at a radius $\mathcal{R}$), then the two conditions can be met only if 
the normal component of $\mathbf{B}$ vanishes.
In spherical coordinates, the boundary conditions at the shell are
therefore, $B_r = 0, [B_\theta]=0$ and $[B_\phi]=0$.
This condition on $B_r$ at $r=\mathcal{R}$ can be met only if
\begin{equation}
 g_{m+3/2}(\alpha r)|_{r=\mathcal{R}}=0.
\end{equation}
 Let 
\begin{equation}
g_{m+3/2}(\alpha r)=c_1 J_{m+3/2}(\alpha r)+c_2 Y_{m+3/2}(\alpha r)
\end{equation}
where $c_1$ and $c_2$ are constants to be determined from the boundary conditions.
Finiteness of $g$ at $r=0$ demands $c_2=0$; physically this implies that the poloidal flux, is finite. Then
\begin{equation}
g_{m+3/2}(\alpha r)=c_1 J_{m+3/2}(\alpha r).
\label{far}
\end{equation}
Finally, the expression for magnetic field is given by
\begin{equation}
\mathbf{B}=\left(-\frac{1}{r^2}\frac{\partial}{\partial\mu}[S_mr^2(1-\mu^2)],\frac{-1}{r\sqrt{(1-\mu^2)}}\frac{\partial}{\partial r}[S_mr^2(1-\mu^2)],
\alpha r\sqrt{(1-\mu^2)}S_m\right).
\label{bchand1}
\end{equation}
The above expression can be further simplified by substituting for $S_m$ using 
Equations (\ref{sn}) and (\ref{far}) to
\begin{eqnarray}
\mathbf{B}&=&\Bigl(\frac{-J_{m+3/2}(\alpha r)}{r^{3/2}}\frac{d }
{d \mu}[(1-\mu^2)C_m^{3/2}(\mu)],\nonumber\\
&&\frac{-1}{r}\frac{d }
{d r}[r^{1/2}J_{m+3/2}(\alpha r)](1-\mu^2)^{1/2}C_m^{3/2}(\mu),
\frac{\alpha J_{m+3/2}(\alpha r)}{r^{1/2}}(1-\mu^2)^{1/2}C_m^{3/2}(\mu)\Bigr).
\label{bchand}
\end{eqnarray}
\normalsize
The various modes of C modes are shown in Figure\ref{cplt} for different values of the variable $m$.
Note that $m$ represents the number of angular oscillations of the mode.
 The total number of poles in the sphere are $2m$.
The self-similarity of the solutions is evident from Figure \ref{cplt}.

 The derivation for the potential field corresponding to 
Equation (\ref{bchand})is given in Appendix \ref{pot}. The final expression
for the potential field is found to be 
\begin{equation}
\mathbf{B}_P=\left(\left[(m+1) a_{m+1} r^{m}-\frac{(m+2)b_{m+1}}{r^{m+3}}\right]P_{m+1}(\mu),-(1-\mu^2)^{1/2}
\left[ a_{m+1} r^{m}+\frac{b_{m+1}}{r^{m+3}}\right]\frac{dP_{m+1}}{d\mu},0\right),
\label{bpchand1}
\end{equation}
where $P_{(m+1)}(\mu)$ are the Legendre polynomials, and where the coefficients are given as
\begin{eqnarray}
\chi_l=\chi_{m+1}(r_1)&=&\frac{(m+1)(m+2)}{r_1^{3/2}}J(m+3/2,\alpha r_1)\\
a_l=a_{m+1}&=&\frac{\chi_{m+1}(r_1)}{(m+1)}\frac{r_1^{m+3}}{ r_1^{2m+3}-r_2^{2m+3}}\nonumber\\
b_l=b_{m+1}&=&\frac{(m+1)}{(m+2)}a_{m+1} r_2^{(2m+3)}
\end{eqnarray}
The above expressions can be further simplified by substituting for $\chi_l$, which gives
\begin{equation}
a_{m+1}=\frac{(m+2)r_1^{m+3/2} J_{m+3/2}(\alpha r_1)}{ r_1^{2m+3}-r_2^{2m+3}};\quad
b_{m+1}=\frac{(m+1)r_2^{2m+3} r_1^{m+3/2} J_{m+3/2}(\alpha r_1)}{ r_1^{2m+3}-r_2^{2m+3}}
\label{ambm}
\end{equation}
\section{\uppercase{Energy for closed field lines of C mode}}
\label{enrg}
\begin{figure}[ht!]
\centerline{\includegraphics[scale=0.25]{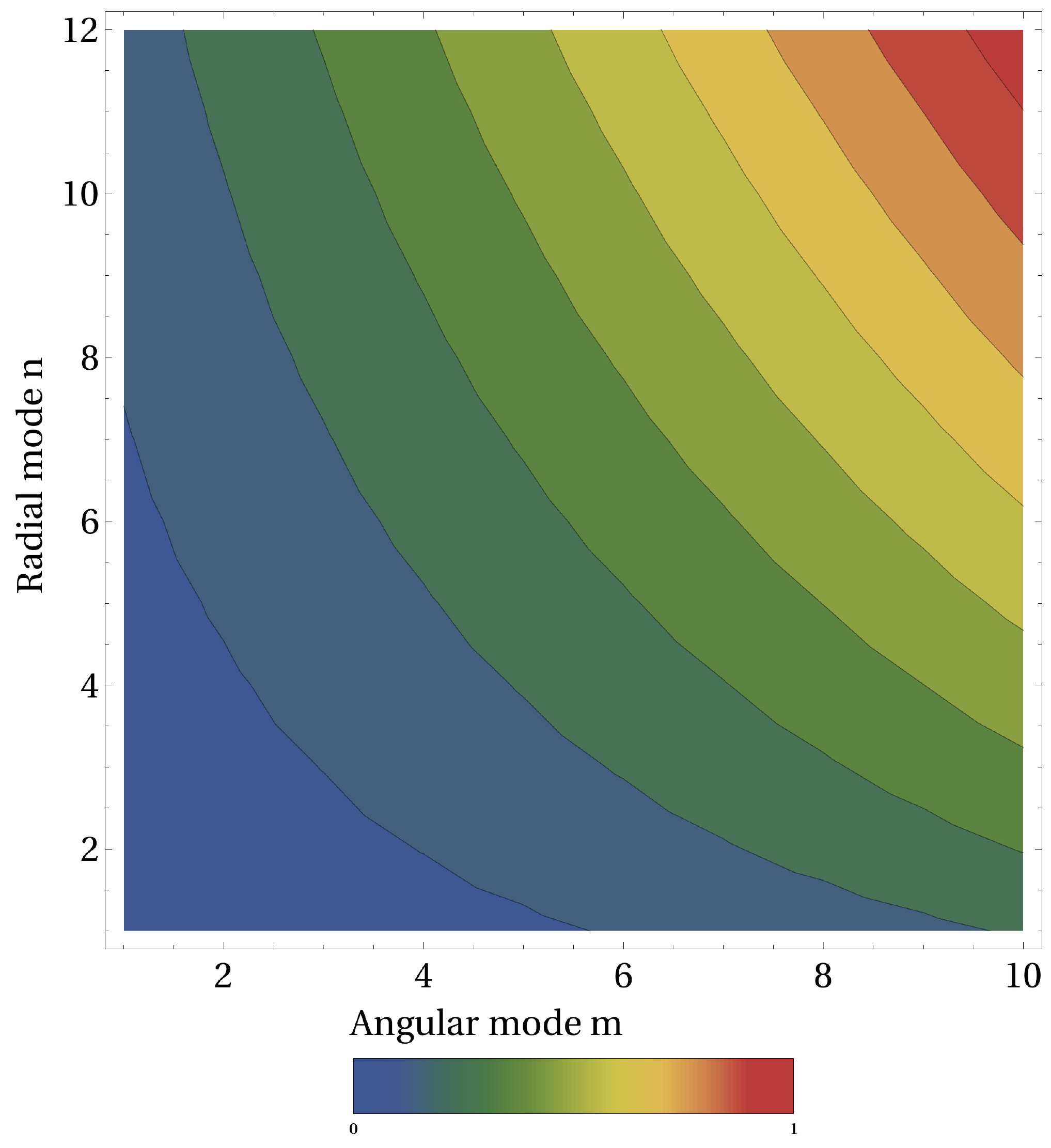}\includegraphics[scale=0.5]{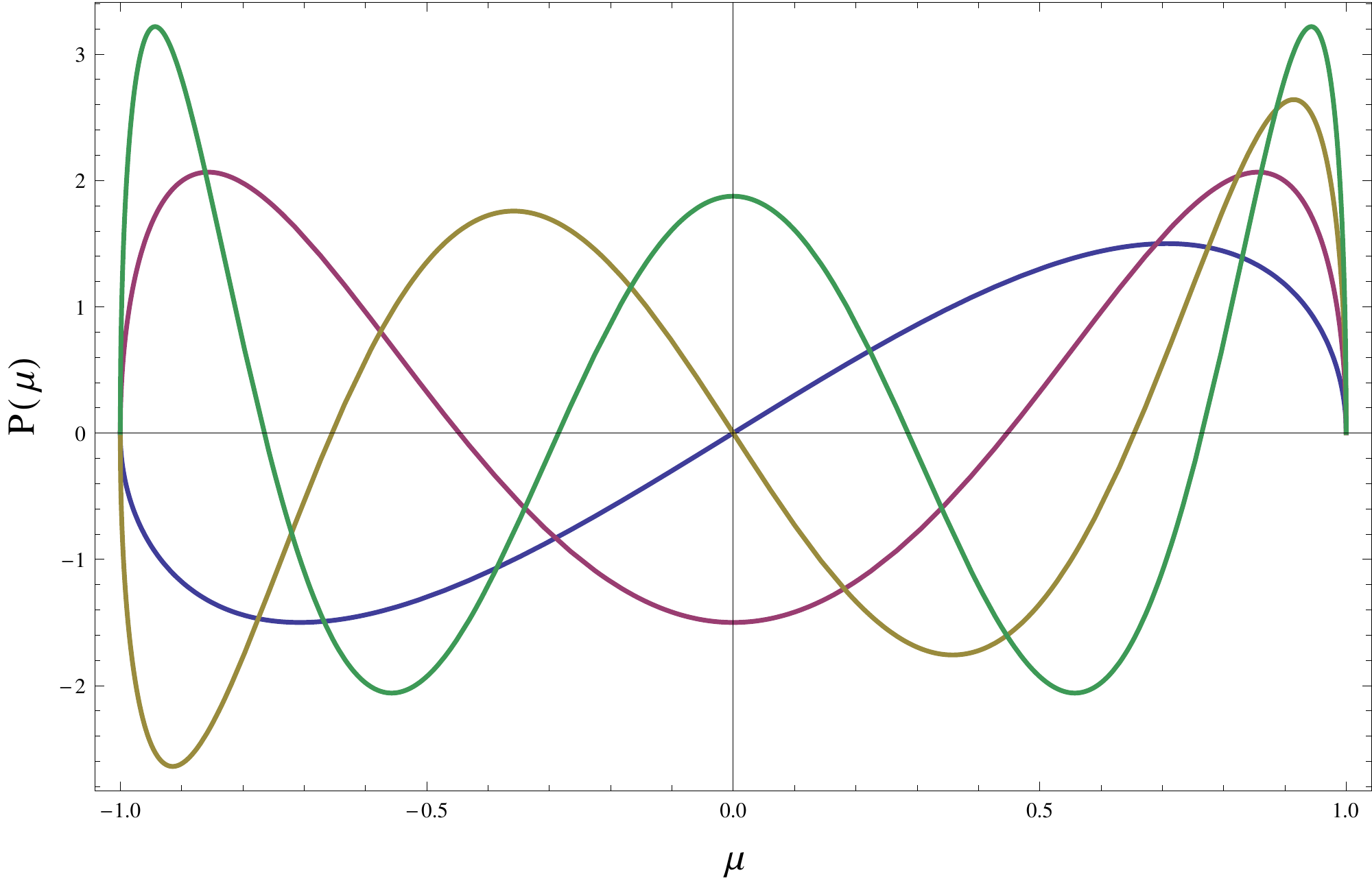} }
\caption{Left panel shows contours of energy for different angular and radial modes for C modes
computed for the same volume and normalized with respect to the maximum. 
The 0 and 1 in the legend refer to the maximum and minimum values of the energies respectively.
The right panel shows the behavior of $P(\mu)$, which changes sign $(m+1)$ times in the domain
for a given value of $m$.}
\label{enm}
\end{figure}

The energy of a force-free magnetic field in a spherical shell geometry is given by
\begin{equation}
E(\mathbf{B})=\frac{1}{8\pi}\int_{r_1}^{r_2}\int_{-1}^1\int_0^{2\pi}|\mathbf{B}|^2r^2 dr d\mu d\phi
=\frac{1}{4}\int_{r_1}^{r_2}\int_{-1}^1|\mathbf{B}|^2r^2 dr d\mu\label{eff1},
\end{equation}
where axisymmetry is applied in the last step.  
The expression for energy for the force-free field given in Equation (\ref{eff1}) uses a 
volume integral, whereas we can calculate it via a surface integral using the virial theorem for force-free fields (Chandrasekhar 1961) in spherical geometry as
\begin{equation}
 E_{\mathrm{ff}}=\frac{1}{8 \pi}\int_V |\mathbf{B}|^2dV= \frac{1}{8 \pi}\int_S|\mathbf{B}|^2\mathbf{r}\cdot d\mathbf{S}
-\frac{1}{4 \pi}\int_S (\mathbf{B}\cdot\mathbf{r})(\mathbf{B}\cdot d\mathbf{S})
\label{eviri}
\end{equation}
where $S$ is the surface enclosing the volume of interest $V$. In axisymmetry, the Equation (\ref{eviri}) reduces to 
\begin{equation}
 E_{\mathrm{ff}}= \frac{1}{8 \pi}\int_V|\mathbf{B}|^2 dV=\frac{ \mathcal{R}^3}{4}\int_{-1}^1(B_\theta^2+B_\phi^2-B_r^2)d\mu.
\label{vir3}
\end{equation}
where $\mathcal{R}$ is the radius of the shell. For the energy of a  potential field, 
we use $E_P=E(\mathbf{B}_P)$, which is calculated from Equation (\ref{eff1}).
In order to study the dependence of energy on the various radial and angular modes,
we first calculate the contribution from the toroidal component given by
\begin{equation}
 E_T(\alpha,n,m,r_1,r_2) =\frac{1}{4}\alpha^2\int_{r_1}^{r_2} dr~ r J^2_{m+3/2}(\alpha r)\int_{-1}^1 d\mu (1-\mu^2)\left[C_m^{3/2}(\mu)\right]^2.
\label{etor}
\end{equation}
using Equation (\ref{eff1}) and $E(\mathbf{B}) =E(\mathbf{B}_T)+E(\mathbf{B}_P)$; the total energy
(for volumes containing closed-field lines) is given by $E_{\mathrm{ff}}=2 E_T$ \citep{chandra61}. 
The energy for C mode can be calculated analytically if the field lines close at the inner and 
outer boundaries. The radial part of the integration in Equation (\ref{etor}) can be written as
\begin{equation}
R=\alpha^2\int_{r_1}^{r_2} dr~ r J^2_{m+3/2}(\alpha r).
\label{rin}
\end{equation}
If $r_1=0$ and $r_2=r_{nm}$, where $Z_{nm}=\alpha r_{nm}$ is the $n$th root of $J_{m+3/2}(\alpha r)$, then Equation (\ref{rin}) can be written as
\begin{equation}
 R=\frac{1}{2} (Z_{nm})^2\left[J_{m+5/2}(Z_{nm})\right]^2.
\label{rpart}
\end{equation}
The angular part of Equation (\ref{etor}) can be written as
\begin{equation}
 \Theta=\int_{-1}^1d\mu (1-\mu^2) C_m^{3/2}(\mu)^2.
\end{equation}
Using the orthogonality properties of Gegenbauer polynomials, the above integral can be evaluated as
\begin{equation}
 \Theta=\frac{\pi \Gamma(m+3)}{4!(m+3/2)[\Gamma(3/2)]^2}=\frac{2(m+1)(m+2)}{2m+3}.
\label{tpart}
\end{equation}
Combining Equations (\ref{rpart}) and (\ref{tpart}), 
we obtain the following expression for Equation (\ref{etor})
\begin{equation}
 E_T=\frac{Z_{nm}^2}{4}\frac{(m+1)(m+2)}{2m+3}\left[J_{m+5/2}(Z_{nm})\right]^2.
\end{equation}
 and plot contours of the result in Figure \ref{enm} (left panel). 
We find that for a sphere of fixed radius, the energy increases
 with higher angular $m$ and radial $n$ modes.
The radial modes of the solution are given by the
Bessel functions, which represent the number of
 radial oscillations and the energy of the field increases with the 
number of oscillations. The angular modes are
 given by $(1-\mu)^{1/2}C_m^{3/2}(\mu)$, which are presented in the right 
panel of Figure \ref{enm}, and the field reverses $(m+1)$ times for a given value of $m$.

\section{\uppercase{Matching potential fields to force-free fields at the inner shell}}
\label{pot}
A potential field is defined by the equation
\begin{equation}
 \mathbf{\nabla}\times\mathbf{B}_P=0.
\end{equation}
Thus the field can be expressed as $\mathbf{B}_P=\mathbf{\nabla}\Phi_P$
for a scalar potential $\Phi_P$ which satisfies the Laplace equation
$\mathbf{\nabla}^2\Phi_P=0\label{potphi}$.
The general solution for this equation in spherical coordinates is given by
 \begin{equation}
\Phi_P(r,\mu)=\sum_{l=0}^{\infty}(a_l r^l+\frac{b_l}{r^{l+1}})P_l(\mu)\label{phisol}
\end{equation}
where $P_l(\mu)$ is the Legendre polynomial of order $l$; $a_l$ and $b_l$ are constant
 coefficients to be determined by matching the normal components of potential field with that of 
the force-free field at the boundaries. We have to solve the Laplace equation for a spherical shell
with $r_1$ and $r_2$ as inner and outer boundaries.

The radial component of the potential field  $[B_r(r,\mu)]_P$is given by
\begin{equation}
 [B_r(r,\mu)]_P=(\mathbf{\nabla}\Phi)_r=\frac{\partial\Phi}{\partial r}
=\sum_{l=0}^{\infty} P_l(\mu)\chi_l(r)\label{brpot}
\end{equation}
where $\chi_l$ is given by
\begin{equation}
{\chi_l(r)=\left[l a_l r^{l-1}-(l+1)\frac{b_l}{r^{l+2}}\right]}.\label{clr}
\end{equation}
We assume that the radial component of the force-free magnetic field can be separated as functions of $r$ and $\mu$
denoted by $R(r)$ and $\Theta(\mu)$, respectively:
\begin{equation}
 \left[B_r(r,\mu)\right]_{ff}=R(r)\Theta(\mu).\label{brrt}
\end{equation}
To match the radial components of potential and force-free fields,
we use equations (\ref{brpot}) and (\ref{brrt}) and equate
 the two fields at the lower boundary, $r=r_1$:
\begin{equation}
R(r_1)\Theta(\mu)=\sum_{l=0}^{\infty} P_l(\mu)\chi_l(r_1).
\end{equation}
Using the orthogonality property of Legendre functions we get
\begin{equation}
R(r_1)\int_{-1}^1\Theta(\mu)P_l(\mu) d\mu=\frac{2}{(2l+1)}\chi_l(r_1).\label{orth}
\end{equation}
So, the expansion coefficients for the potential field can be obtained from Equation (\ref{orth}) as
\begin{equation}
\chi_l(r_1)=\frac{(2l+1)}{2}R(r_1)\int_{-1}^1\Theta(\mu)P_l(\mu) d\mu
\label{clr1}
\end{equation}
\subsection{Matching Potential Field to C Modes at the Inner Shell}
\label{cpot}
Using Equations (\ref{clr1}) and (\ref{far}) we can write at the inner boundary $r=r_1$
\begin{eqnarray}
\chi_l(r_1)&=&\frac{-(2l+1)}{2 r_1^{3/2}}J(m+3/2,\alpha r_1)\int_{-1}^1\frac{\partial}{\partial \mu}\left[C^{3/2}_m(\mu)(1-\mu^2)\right]P_l(\mu) d\mu
\label{muin} \\
&=&\frac{(m+1)(m+2)}{r_1^{3/2}}J(m+3/2,\alpha r_1),
\label{cl} 
\end{eqnarray}
where $l=m+1$; the calculation of the $\mu$ integral in Equation (\ref{muin}) is given in appendix section \ref{muap}.
At the outer boundary at $r=r_2$, we have $\chi_l(r_2)=0$, which results in the following condition for the coefficients $a_l$ and $b_l$
\begin{equation}
b_l= b_{(m+1)}=\frac{(m+1)}{(m+2)}a_{(m+1)} r_2^{(2m+3)}.\label{bl}
\end{equation}
Using  Equation (\ref{clr}) and the above equation, we find the following expression for the coefficient $a_{(m+1)}$
\begin{equation}
a_l= a_{(m+1)}=\frac{\chi_{(m+1)}(r_1)}{(m+1)}\frac{r_1^{(m+3)}}{ r_1^{(2m+3)}-r_2^{(2m+3)}}.\label{al}
\end{equation}
Upon simplification the coefficients can be written as

\begin{equation}
a_{m+1}=\frac{(m+2)r_1^{m+3/2} J_{m+3/2}(\alpha r_1)}{ r_1^{2m+3}-r_2^{2m+3}};\quad
b_{m+1}=\frac{(m+1)r_2^{2m+3} r_1^{m+3/2} J_{m+3/2}(\alpha r_1)}{ r_1^{2m+3}-r_2^{2m+3}}
\end{equation}
Thus the expression for the potential field is given by
\begin{equation}
\mathbf{B}_P=\left(\left[(m+1) a_{(m+1)} r^{m}-\frac{(m+2)b_{(m+1)}}{r^{(m+3)}}\right]P_{(m+1)}(\mu),
-(1-\mu^2)^{1/2}\left[ a_{(m+1)} r^{m}+\frac{b_{(m+1)}}{r^{(m+3)}}\right]\frac{dP_{(m+1)}}{d\mu},0\right).
\label{Bpchand}
\end{equation}

\subsection{Matching Potential Field to LL Modes at the Inner Shell}
\label{llpot}
We recall the definitions for the general potential field from Equation (\ref{phisol}).
Now the boundary condition at the outer boundary $r_2 (=\infty)$ is given by
$\chi_l(r_2)=0$, and because the potential should be finite for all values of $r$, it implies that
\begin{equation} 
a_l=0
\label{alLL}
\end{equation}
and the scalar potential takes the form
\begin{equation}
 \Phi_P(r,\mu)=\sum_{l=0}^{\infty}\frac{b_l}{r^{l+1}}P_l(\mu)
\label{nlphi}
\end{equation}
whereas the radial component of the potential field is given by
\begin{equation}
 B_r(r,\mu)=\sum_{l=0}^{\infty}-(l+1)\frac{b_l}{r^{l+2}}P_l(\mu).
\end{equation}
From Equation (\ref{low}) we recall that the radial component
 of the nonlinear field has the following form 
\begin{equation}
 B_r(r,\mu)=-\frac{1}{r^{n+2}}\frac{d P}{d\mu}.
\end{equation}
Equating the two radial fields at the lower boundary, $r=r_1$ we get
\begin{equation}
\sum_{l=0}^{\infty}-(l+1)\frac{b_l}{r_1^{l+2}}P_l(\mu) =-\frac{1}{r_1^{n+2}}\frac{dP}{d\mu}.
\label{pplcond}
\end{equation}
Using the orthogonality property of the Legendre polynomials
we get the following expression for the expansion coefficient $b_l$:
\begin{equation}
b_l=\frac{2l+1}{2(l+1)}
r_1^{l-n}\int_{-1}^1\frac{dP}{d\mu}P_l(\mu)d\mu.
\label{blLL}
\end{equation}
So the final expression for the potential field matched to LL modes is given by
\begin{equation}
\mathbf{B}_P=\left(\sum_{l=0}^{\infty}-(l+1)\frac{b_l}{r^{l+2}}P_l(\mu),\sum_{l=0}^{\infty}\frac{-b_l}{r^{l+2}}
(1-\mu^2)^{1/2}\frac{dP_l}{d\mu},0\right).
\label{Bplow}
\end{equation}

\section{\uppercase{Vector potential of potential fields}}
\label{vpot}
The vector potential for the potential field is given by the relation
\begin{equation}
 \mathbf{\nabla}\times\mathbf{A}_P=\mathbf{\nabla}\Phi_P,
\end{equation}
where $\Phi_P$ is the scalar potential obtained from Equation (\ref{phisol}).
Because a potential field is entirely poloidal and the 
curl of a toroidal field is always poloidal, 
we expect $\mathbf{A}_P$ to have only toroidal components.
Then an axisymmetric field $\mathbf{A}_P$ will be of the following form:
\begin{equation}
 \mathbf{A}_P=A_\phi(r,\mu)\hat{\phi}
\end{equation}
Expanding the above equation in spherical polar coordinates, we obtain the following pair of equations
\begin{eqnarray}
&& \frac{-1}{r}\frac{\partial}{\partial \mu}\left[(1-\mu^2)^{1/2}A_\phi\right]=\frac{\partial \Phi_P}{\partial r}\label{vpot1}\\
&&\frac{1}{r}\frac{\partial}{\partial r}(r A_\phi)=\frac{(1-\mu^2)^{1/2}}{r}\frac{\partial \Phi_P}{\partial \mu}\label{vpot2}
\end{eqnarray}
Solving the above two equations simultaneously, we find the unique solution
\begin{equation}
A_\phi(r,\mu)=\sum_{l=0}^{\infty}(1-\mu^2)^{1/2} P^\prime_l(\mu)\left[\frac{ a_l r^l}{l+1}-\frac{b_l}{lr^{l+1}}\right].
\end{equation}
\begin{equation}
A_\phi(r,\mu)=\sum_{l=0}^{\infty}(1-\mu^2)^{1/2} P^\prime_l(\mu)\left[\frac{ a_l r^l}{l+1}-\frac{b_l}{lr^{l+1}}\right].
\end{equation}
So, the final expression is given by
\begin{equation}
\mathbf{A}_P=\sum_{l=0}^{\infty}\left(0,0,(1-\mu^2)^{1/2} P^\prime_l(\mu)\left[\frac{ a_l r^l}{l+1}-\frac{b_l}{lr^{l+1}}\right]\right).
\label{Apchand}
\end{equation}
\section{\uppercase{Vector potential for LL modes}}
\label{llvpot}
Because $\mathbf{A}$ is uncertain within a choice of gauge, we choose a convenient gauge such that the radial component
of the vector potential, $A_r$ is zero. Then the vector potential in spherical polar coordinates can be written as
\begin{equation}
 \mathbf{A}=(0,A_\theta,A_\phi).
\end{equation}
Using the definition $\mathbf{B}=\mathbf{\nabla}\times\mathbf{A}$, 
we get the following three equations for the components of $\mathbf{A}$
\begin{eqnarray}
 \frac{-1}{r^2}\frac{\partial \psi}{\partial \mu}&=&\frac{-1}{r}\frac{\partial }{\partial \mu}\left[(1-\mu^2)^{1/2}A_\phi\right]\nonumber\\
 \frac{-1}{r(1-\mu^2)^{1/2}}\frac{\partial \psi}{\partial r}&=&\frac{-1}{r}\frac{\partial }{\partial r}(r A_\phi)\nonumber\\
\frac{a \psi^{(n+1)/n}}{(1-\mu^2)^{1/2}}&=&\frac{\partial (r A_\theta)}{\partial r}.
\end{eqnarray}
By solving the above set of equations, we find
\begin{equation}
 \mathbf{A}=\left(0,\frac{-a}{n r^{n+1}}\frac{P(\mu)^{(n+1)/n}}{(1-\mu^2)^{1/2}},\frac{1}{r^{n+1}}\frac{P(\mu)}{(1-\mu^2)^{1/2}}\right)
\label{alow}
\end{equation}
As a consequence of equations (\ref{low}, \ref{alow}), $\mathbf{A}\cdot\mathbf{B}=0$ everywhere. 
For closed-field lines in a volume, the magnetic helicity
\begin{equation}
 H=\int\mathbf{A}\cdot\mathbf{B}\mathrm~{d}V=0.
\end{equation}

\section{\uppercase{Equivalence of Finn Antonsen and Berger formulae for force-free spheres}}

\subsection{C Modes}
\label{fabc}

To show the equivalence of expressions of relative helicity obtained 
from Equations (\ref{hrelchand}) and (\ref{hrelcb}), 
we first express Equation (\ref{hrelchand}) as
\begin{equation}
 H_r=\frac{8\pi E_{\mathrm{ff}}}{\alpha}+\frac{4\pi(m+1)(m+2)}{\alpha(2m+3)}(I_1+I_2)
\label{hri12}
\end{equation}
where $E_{\mathrm{ff}}$ is given by Equation (\ref{eint}) and $I_1$ is the integral given by
\begin{equation}
I_1=\int_{r_1}^{r_2} \alpha^2 \left(\frac{a_{m+1}r^{m+1}}{m+2}-
\frac{b_{m+1}}{(m+1)r^{m+2}}\right) r^{3/2}J_{m+3/2}(\alpha r)
\end{equation}
and $I_2$ is the boundary term given by
\begin{equation}
 I_2=r_1^{1/2}\left(a_{m+1}r_1^{m+1}+\frac{b_{m+1}}{r_1^{m+2}}\right)J_{m+3/2}(\alpha r_1).
\end{equation}
Upon simplification, we get 
\begin{equation}
 I_1+I_2=\left(\frac{a_{m+1}r_1^{m+2}}{m+2}-\frac{b_{m+1}}{(m+1)r_1^{m+1}}\right)
\frac{d}{dr}\left[r^{1/2}J_{m+3/2}(\alpha r)\right]|_{r=r_1}.
\end{equation}
Now from the continuity of the radial component of the force-free field 
to the potential field  at $r=r_1$,
$(\mathbf{B}_P)_r=(\mathbf{B})_r$, where $\mathbf{B}$ and $\mathbf{B}_P$ are 
given by Equations (\ref{bchand}) and (\ref{bpchand1}) respectively, we find
\begin{equation}
 \left(\frac{a_{m+1}r_1^{m+2}}{m+2}-\frac{b_{m+1}}{(m+1)r_1^{m+1}}\right)=
 r_1^{1/2}J_{m+3/2}(\alpha r_1)
\end{equation}
which leads to
\begin{equation}
 I_1+I_2=r_1^{1/2}J_{m+3/2}(\alpha r_1)
\frac{d}{dr}\left[r^{1/2}J_{m+3/2}(\alpha r)\right]|_{r=r_1}.
\label{i12}
\end{equation}
Substituting Equations (\ref{eint}) and (\ref{i12}) 
in Equation (\ref{hri12}) we arrive at Equation (\ref{hrelcb}).

\subsection{LL Modes}
\label{fabll}

To prove the equivalence of the expressions of relative helicity given in Equations (\ref{hrellow} \& \ref{hrbll}),
we start with Equation (\ref{pplcond}), which can be rewritten as

\begin{equation}
\sum_{l=0}^{\infty}(l+1)\frac{b_l}{r_1^{l}}P_l(\mu) =\frac{1}{r_1^{n}}\frac{dP}{d\mu}.
\label{pplcond1}
\end{equation}
Integrating the above equation with respect to $\mu$ and rearranging the terms, we get
\begin{equation}
 \sum_{l=0}^{\infty}(l+1)\frac{b_l}{r_1^{l-n}}\int P_l(\mu) d\mu =P.
\label{lleq1}
\end{equation}
Now from the Legendre differential equation, we have the identity
\begin{equation}
 \left[(1-\mu^2)\frac{dP_l}{d\mu} \right ]=-l(l+1)\int P_l d\mu.
\label{lleq2}
\end{equation}
Substituting Equation (\ref{lleq2}) in Equation (\ref{lleq1}), we get
\begin{equation}
-\sum_{l=0}^{\infty}\frac{b_l}{l r_1^{l-n}}\left[(1-\mu^2)\frac{dP_l}{d\mu} \right ]= P.
\label{lleq3}
\end{equation}
Multiplying both sides of Equation (\ref{lleq3})
 by $\displaystyle{\frac{2\pi a P^{1+1/n}}{n r_1^{2n}(1-\mu^2)}}$,
we get the equality between the integrands of Equations (\ref{hrellow}) and (\ref{hrbll}).

\section{\uppercase{Calculation of angular integral in the expression} (\ref{muin})}
\label{muap}
Here we give the derivation of the angular integral in Equation (\ref{muin}):
\begin{equation}
 \int_{-1}^1\frac{\partial}{\partial \mu}\left[C^{3/2}_m(\mu)(1-\mu^2)\right]P_l(\mu) d\mu.\label{min}
\end{equation}
We now expand Gegenbauer polynomials in terms of Legendre
 polynomials by using the following relation:
\begin{equation}
 (1-\mu^2) C_m^{3/2}(\mu)= (1+m)\left[P_m(\mu)-\mu P_{m+1}(\mu)\right]=(1-\mu^2)P_{m+1}^\prime (\mu).
\end{equation}
Equation (\ref{min}) can now be written as
\begin{eqnarray}
&&\int_{-1}^1 \frac{\partial}{\partial \mu}\left[(1-\mu^2)P_{m+1}^\prime (\mu)\right] P_l(\mu)d\mu\nonumber\\
&=&\left[ P_l(\mu)(1-\mu^2)P_{m+1}^\prime (\mu)\right]_{-1}^1-\int_{-1}^1 P_l^\prime (\mu) P_{m+1}^\prime (\mu)(1-\mu^2)d\mu\nonumber\\
&=& \int_{-1}^1 P_l^\prime(\mu)d\mu\int(m+1)(m+2)P_{m+1}(\mu)d\mu\nonumber\\
&=& \left[\int(m+1)(m+2)P_{m+1}(\mu)~P_l(\mu)~d\mu\right]_{-1}^1-\int(m+1)(m+2)P_{m+1}(\mu)P_l(\mu)d\mu\nonumber\\
&=& -(m+1)(m+2)\delta_{0,m+1}P_l(\mu)-\frac{2(m+1)(m+2)}{2l+1}\delta_{l,m+1}\nonumber\\
&=&-\frac{2(m+1)(m+2)}{2m+3},
\end{eqnarray}
where we have used Legendre differential equations to substitute for the derivative of $P^\prime_{m+1}(\mu)$ in the third step and
the orthogonality property of Legendre polynomials in the final step.

\section{\uppercase{Boundary conditions for $F$ in Equation} (\ref{feq})}
\label{fbcond}
We motivate the transformation of the variable from $P$ to $F$ by $P=(1-\mu^2)^{1/2}F$.
This enables us to write an ODE Equation (\ref{feq}) to solve directly for LL fields for all of the 
allowed cases of $n$ that are numerically difficult to implement with the ODE for $P$, given 
by Equation (\ref{peq}).
The angular part of the LL mode is  given by Equation (\ref{peq})
\begin{equation}
(1-\mu^2)P^{\prime\prime}+a^2\frac{n+1}{n}P^{1+2/n}+n(n+1)P=0.
\end{equation}
We assume
\begin{equation}
 P(\mu)= (1-\mu^2)^\Gamma g(\mu),
\label{pgam}
\end{equation}
as $P(\mu=\pm 1)=0$ for the highest possible $\Gamma >0$ such that $g(\mu=\pm 1) \neq 0$.
Substituting for $P$  in Equation (\ref{peq}) we obtain
\begin{eqnarray}
 (1-\mu^2)^2 g^{\prime\prime}+\left[-2 \Gamma (1-\mu^2) + 4\mu^2 \Gamma(\Gamma-1) + n(n+1)(1-\mu^2)\right]g && 
\nonumber\\- 4\mu\Gamma(1-\mu^2)g^\prime + a^2\frac{(n+1)}{n}g^{\frac{n+2}{n}}(1-\mu^2)^{\frac{2\Gamma}{n}+1}&=&0. 
\label{ggamma}
\end{eqnarray}
We now expand $P$ using Equation (\ref{pgam}) in a power series of $(1-\mu^2)$ near $\mu = \pm 1$,
\begin{equation}
 P(\mu)=(1-\mu^2)^\Gamma\sum_{\gamma=0}^\infty  C_\gamma (1-\mu^2)^\gamma,
\label{pexp}
\end{equation}
where $C_0$ is the leading term which is nonzero by definition.
Comparing Equation (\ref{pgam}) and (\ref{pexp}), we can expand $g(\mu)$ near $\mu^2=1$ 
in a power series with coefficients $C_\gamma$ as
\begin{equation}
 \lim_{\mu^2 \to 1}g(\mu)= C_0+C_1 (1-\mu^2)+ C_2(1-\mu^2)^2+... 
\end{equation}
It is clear from above that in the limit $\mu^2\rightarrow1$, $g\rightarrow C_0$ 
which is a constant.
Also we know that $g^\prime$ and $g^{\prime\prime}$ are 
finite as $\mu^2\rightarrow1$ because $P(\mu)$
is finite in this limit. As a result, upon the substitution $\mu^2\rightarrow1$, 
Equation (\ref{ggamma}) gives
\begin{equation}
 4\mu^2 \Gamma(\Gamma-1) g=0,
\end{equation}
leading to $\Gamma=0, 1$. The $\Gamma=0$ solution is not allowed whereas $\Gamma=1$ implies
\begin{equation}
 P(\mu)= (1-\mu^2) g = (1-\mu^2)^{1/2} F.
\end{equation}
Thus $F$ satisfies the boundary conditions
\begin{equation}
 F(\mu)=0\quad \textrm{at}\quad \mu=-1,1.
\end{equation}

\end{document}